\documentclass[11pt,a4paper]{article}
\usepackage[T1]{fontenc}
\usepackage[utf8]{inputenc}
\usepackage[british]{babel}
\usepackage{amsmath,amssymb,amsfonts,bm}
\usepackage[margin=2.4cm]{geometry}
\usepackage{graphicx,float}
\usepackage{xcolor}
\definecolor{linkcol}{RGB}{30,70,120}
\usepackage[colorlinks=true,allcolors=linkcol]{hyperref}
\newcommand{\K}{\mathbf{K}}
\newcommand{\R}{\mathbf{R}}
\newcommand{\Mc}{\mathcal{M}}
\newcommand{\Lc}{\mathcal{L}}
\newcommand{\avg}[1]{\left\langle #1 \right\rangle}
\newcommand{\Keff}{K a\cos\tilde\beta}

\title{\textbf{Surface diffusion with lateral interactions: a closed-form
intermediate scattering function of an Ising adlayer from the memory-equation
formalism}}
\author{S. Miret-Art\'es\\[4pt]
\small Instituto de F\'isica Fundamental, CSIC, Serrano 123, 28006 Madrid, Spain}
\date{\today}

\begin{document}
\maketitle

\begin{abstract}
\noindent
The intermediate scattering function (ISF) 
and its value at zero time, which is the static structure factor
(SSF), are both characteristic functions (CF). We use this double role to bring lateral 
interactions of Ising lattice-gas type into surface diffusion. 
An exact equation of motion for the ISF carrying a local rate and a memory function
is then obtained. To interpret this equation, the ISF is expressed in terms of relaxation
modes. The local rate is fixed
by a sum rule which reversibility turns into an equilibrium average. 
The  de Gennes narrowing  is established as a theorem. 
The memory function being non-negative, the closed form that results is a
rigorous lower bound on the ISF at every momentum transfer and every time, and
its accuracy is itself predicted.
In the diffusive regime, the closed form
keeps the single-exponential shape to which spin-echo data are routinely fitted,
with both coefficients renormalized by the interaction: the amplitude becomes
the SSF of the layer and the dephasing rate the dressed one-adsorbate rate
divided by it, explicit in momentum transfer, time, coverage and temperature.
The whole hierarchy of linear-response functions is then easily built. 
When introducing the lateral interaction, a single correlation
parameter $\eta$ fixes the equilibrium structure and the renormalized hop
rate: exactly along a one-dimensional channel, and on the surface as the
nearest-neighbour (NN) correlation of a pair approximation completed by a
collective (random-phase) structure factor. The result contains no adjustable parameter and is exact at vanishing
interaction for every coverage and momentum; the Darken relation comes with it, the correlation between successive moves
separated off as a computable factor. 
The interaction leaves three correlated fingerprints --- in the diffusive
amplitude, in its linewidth, and in the Arrhenius slope --- and together they
overdetermine the attempt frequency, the static barrier and the interaction
strength, while the discarded memory remains measurable in the height of the
diffusive peak and in the dependence of the fitted rate on the width of the time window. 
Applications to Na/Cu(111) and H/Pt(111) are
analysed and discussed.
\end{abstract}

\vspace{2pt}

\section{Introduction}
\label{sec:intro}

Surface diffusion is the elementary step behind processes from heterogeneous catalysis to epitaxial growth. Helium atom scattering probes it with atomic-scale resolution, and helium spin-echo (HeSE) does so in the time domain \cite{Jardine2009}. Both are fully coherent and both return the ISF, $I(\mathbf{K},t)$, which is not merely a convenient observable \cite{TorresMiyares2026a,TorresMiyares2026b,TorresMiyares2026c}: it is a CF in the strict sense of probability theory, the Fourier transform of the distribution of adsorbate displacements. Hence its organizing power: it generates the moments of the displacement, and with them the diffusion coefficient; its exponential decay in the diffusive regime follows from the compound-Poisson structure of Markovian jump processes \cite{Montroll1965,TorresMiyares2026c} instead of being assumed, as is customary in the scattering literature \cite{Chudley1961,Frenken1992}; and the whole family of linear-response functions descends from it \cite{TorresMiyares2026c}. 

Reversibility adds the structure on which this work is built: the ISF is described throughout in terms of the relaxation modes of the adlayer. The generator of the dynamics being self-adjoint in equilibrium, its spectrum of decay rates is real and non-negative and, away from the Bragg condition, the ISF is a positive superposition of decaying exponentials, one for each mode into which the density fluctuation at momentum transfer parallel to the surface $\mathbf{K}$ decomposes. The projection of Nakajima and Zwanzig \cite{nakajima,zwanzig60} supplies the exact memory equation --- a generalized master equation with a local rate and a retarded kernel --- that this superposition obeys. This equation is our point of departure, and with it comes the means of asking
how far a single exponential reliable.

What the CF framework has so far lacked is the lateral interaction. For small coverages, the interacting-single-adsorbate (ISA) model reduced it to an uncorrelated collisional friction \cite{MartinezCasado2007}, which retains no information on the sign of the interaction, on its strength, or on the short-range correlations it induces --- exactly the information a coverage-dependent experiment provides, and that
measurements on alkali adlayers have long displayed \cite{Ellis2001,Alexandrowicz2006}. Interacting diffusion has been studied microscopically in the lattice-gas tradition \cite{Reed1981,Gomer1990,Uebing1991,AlaNissila2002}, but those results concern mainly the diffusion coefficients, through the thermodynamic factor, and not the time- and momentum-resolved observable itself. What is missing is a treatment that keeps the interaction and still provides the ISF in closed form.

The closed form is not postulated. Its amplitude is an identity, the value of the
ISF at $t=0$, and its exponential shape is both the compound-Poisson structure of
a Markovian jump sequence and the same form to which measured data are fitted.
Its rate, and the accuracy of that shape, come from the memory equation announced above: eliminating from the dynamics every variable of the layer except the one the measurement uses leaves that equation exact, with its local term and its retarded kernel.
The coefficient of the local term is fixed by a sum rule, and reversibility turns
that sum rule into a manifestly positive equilibrium average in which the
many-body content collapses onto the mean exchange current: its momentum
dependence is the single-jump geometry alone, so that the structure of the layer
enters the rate divided, not multiplied. Expressed on the relaxation modes, this coefficient is the mean of the mode
rates, weighted by the share of the signal each mode carries, while the kernel
records their spread about that mean. The narrowing of the line by the inverse structure factor --- the mechanism de Gennes identified in dense fluids \cite{deGennes1959} --- is therefore a theorem on the lattice, neither an analogy carried over from the liquid nor a closure imposed on the kinetics; the Darken relation follows with it. 
What remains is the memory function, and discarding it is the one step the
dynamics requires. That step is controlled, the kernel being non-negative, the closed
form is a rigorous lower bound on the ISF at every momentum transfer and every
time. Its accuracy is itself predicted by the
spread of the mode rates about their mean. When vanishing identically at
zero interaction, the density wave is itself an exact relaxation mode and the symmetric exclusion process \cite{Kutner1981} is accordingly recovered without error.

We bring the Ising lattice gas --- a nearest-neighbour (NN) pair energy written
on the site occupations --- into the CF framework, where it dresses every analytic ingredient of the ISF while preserving its closed form. A single correlation parameter $\eta$ then governs the equilibrium structure and renormalizes the hop rate: exactly, from the transfer matrix along a one-dimensional (1D) channel, and on the full surface as the NN correlation of a pair approximation together with a collective SSF, the two agreeing at weak coupling. We stay throughout in the spatially uniform, correlated regime, which covers the great majority of experiments of this type.
The interaction then leaves three correlated fingerprints, none of them tied to
a phase transition: a structured diffusive amplitude, an exactly anticorrelated distortion of the linewidth, and a shift of the Arrhenius slope with coverage. Measured together they overdetermine the attempt frequency, the static barrier and the interaction strength, which turns the theory into an internal consistency test instead of a one-parameter fit. We test the first two on the strongly repulsive Na/Cu(111) adlayer, where published spin-echo data \cite{Ward2021} already show the anticorrelation, and predict how raising the coverage would shift both signatures. This system is chosen because the lattice-gas description of its structure factor fails on it, so that what survives is precisely the part of the theory under test. For the transport side, we turn to the attractive H/Pt(111) layer, taking the interaction strength as an explicit range and following how the hop rate, the diffusion coefficients and the activation energy respond together.

The paper is organized as follows. Section~\ref{sec:cf} recalls what makes the
ISF a CF and fixes the single-adsorbate ingredients; Section~\ref{sec:eom}
derives the equation of motion it obeys; Section~\ref{sec:coeff} establishes the sum rule
 and the local rate exactly; Section~\ref{sec:modes} develops the ISF as a superposition of relaxation modes
and bounds the error of the Markovian truncation; Section~\ref{sec:lineshape}
carries through the linear-response hierarchy and the line shape;
Section~\ref{sec:model} introduces the lattice gas and reduces its short-range
statics to the single parameter $\eta$, exactly along a
channel; Section~\ref{sec:ssf} turns $\eta$ into the SSF $F(\mathbf{K})$,
exactly in one dimension and at collective level on the surface;
Section~\ref{sec:dyn} evaluates the exchange current and assembles the
closed-form ISF; Section~\ref{sec:transport} then gives the transport
coefficients and Section~\ref{sec:arrhenius} the effective activation energy;
and Section~\ref{sec:predictions} confronts the theory with experiment.
Section~\ref{sec:concl} concludes.

\section{Theory}
\label{sec:theory}

\subsection{The intermediate scattering function as a characteristic function}
\label{sec:cf}

The adsorbates occupy the sites of a Bravais lattice on a periodic substrate,
each site admitting at most one of them, so that the state of the adlayer is
carried by the occupation numbers $n_{\R}\in\{0 \, (empty),1 \, (occupied)\}$, one per lattice point,
whose common mean is the surface coverage, $\avg{n_{\R}}=\theta$. The substrate acts as a
thermal bath~\cite{TorresMiyares2026a,TorresMiyares2026b,TorresMiyares2026c}. The
state of adlayer and substrate together is carried by a density
operator $\hat\rho_{\rm tot}(t)$, and tracing over the substrate degrees of
freedom leaves the reduced density operator of the adlayer alone,
\begin{equation}
\hat\rho_{\rm red}(t) \;=\; {\rm Tr}_{\rm bath}\,\hat\rho_{\rm tot}(t) .
\label{eq:rhored}
\end{equation}

Diffusion is a dynamical process and deals with  relative positions of adsorbates. 
The starting point is the equilibrium layer conditioned on one adsorbate
occupying the origin at $t=0$, 
the equilibrium layer being stationary.
The reduced operator 
$\hat\rho_{\rm red}(t)$ evolves from it. In the
lattice basis $\{\,|\R\rangle\,\}$, its diagonal element
\begin{equation}
\rho_{\R}(t) \;=\; \avg{\,\R\,|\,\hat\rho_{\rm red}(t)\,|\,\R\,}
\label{eq:rhoR}
\end{equation}
is the probability of finding an adsorbate at the lattice point $\R$ at time $t$,
given that one occupied the origin at time zero; the layer being uniform, the
choice of origin is immaterial and the separation alone labels the state. 
Three properties of \eqref{eq:rhoR} are needed below, and all three are exact.
First, $\rho_{\R}(t)\ge0$ at every time, being a diagonal element of a positive
operator. Second, at $t=0$, it is the equilibrium separation distribution,
$\rho_{\R}(0)=P_{\R}$, with $\rho_{\mathbf 0}(0)=1$, an adsorbate being certainly
at zero separation from itself, and $\rho_{\R}(0)\to\theta$ at large separation,
distant sites being uncorrelated. Third, its total weight is conserved,
\begin{equation}
\sum_{\R}\rho_{\R}(t) \;=\; \sum_{\R}\rho_{\R}(0) \qquad\text{at all } t ,
\label{eq:norm}
\end{equation}
the dynamics moving adsorbates between sites without creating or destroying them.
Divided by that constant, $\rho_{\R}(t)$ is a probability distribution on the
lattice at every time, and not only at $t=0$.

The coherent ISF is the lattice Fourier
transform of that distribution,
\begin{equation}
I(\K,t) \;=\; \sum_{\R} e^{i\K\cdot\R}\,\rho_{\R}(t),
\label{eq:cfdef}
\end{equation}
with $\K$  chosen by the experiment geometry with modulus $K$ and 0a
direction fixed. Setting $\K=\mathbf 0$ identifies the
conserved weight of \eqref{eq:norm} as $I(\mathbf 0,t)$, so that the transform
normalized by it is a CF in the strict sense at every time,
that of the lattice vector $\R$ drawn from $\rho_{\R}(t)$. Its value at $t=0$ is
the static structure factor (SSF),
\begin{equation}
I(\K,0) \;=\; F(\K) \;=\; \sum_{\R} P_{\R}\,e^{i\K\cdot\R},
\label{eq:Fdef}
\end{equation}
itself the CF of the equilibrium separation distribution:
one dynamical CF whose initial value is a second, static
one. Only the diagonal of $\hat\rho_{\rm red}$ enters; the coherences of the
reduced density matrix are not needed to build the measured signal.

The phase in \eqref{eq:cfdef} is expressed through the elementary jumps by which
the lattice point is reached,
\begin{equation}
\R(t) \;=\; \sum_j n_j(t)\,\mathbf L_j ,
\label{eq:jumps}
\end{equation}
where $\mathbf L_j$ is the jump vector along the crystallographic direction $j$,
of length $a_j=|\mathbf L_j|$, and $n_j(t)$ the net number of jumps performed
along it, forward minus backward, the sum running over every jump direction the
lattice offers, NN and longer jumps alike. Only the component of
$\R$ along $\K$ enters the scattered phase,
\begin{equation}
\K\cdot\R(t) \;=\; \sum_j n_j(t)\,\K\cdot\mathbf L_j \;=\; K\,L_{\parallel}(t),
\qquad
L_{\parallel}(t) \;=\; \sum_j n_j(t)\,a_j\cos\tilde\beta_j ,
\label{eq:Lpar}
\end{equation}
with $\tilde\beta_j$ the angle between $\mathbf L_j$ and $\K$. A jump direction
orthogonal to $\K$ carries $\cos\tilde\beta_j=0$ and drops out of
\eqref{eq:Lpar} irrespective of the number of jumps made along it, so that $L_{\parallel}$
counts only the directions not orthogonal to the momentum transfer, each weighted
by the cosine of its angle with $\K$. It is a real number, not a lattice vector, and it is the variable the
experiment resolves. Carrying
\eqref{eq:Lpar} into \eqref{eq:cfdef},
\begin{equation}
I_N(\K,t) \;\equiv\; \frac{I(\K,t)}{I(\mathbf 0,t)}
\;=\; \avg{\,e^{\,iKL_{\parallel}(t)}\,},
\label{eq:cfnorm}
\end{equation}
the average being taken over $\rho_{\R}(t)$ normalized by its own total weight:
the ISF is the CF of the projected displacement
$L_{\parallel}$, times a fixed number. It therefore generates the moments of
$L_{\parallel}$ by differentiation with respect to $K$ at fixed direction, and
the cumulants by differentiation of its logarithm,
\begin{equation}
\avg{L_{\parallel}^{\,n}(t)} \;=\; (-i)^n\,
\frac{\partial^{\,n}I_N(\K,t)}{\partial K^{\,n}}\bigg|_{K=0},
\qquad
\kappa_n(t) \;=\; (-i)^n\,
\frac{\partial^{\,n}\ln I_N(\K,t)}{\partial K^{\,n}}\bigg|_{K=0} ,
\label{eq:cumulants}
\end{equation}
The layer being uniform, forward and backward displacements are equally
likely, so every odd order vanishes; in particular $\avg{L_{\parallel}}=0$,
and the second cumulant coincides with the second moment --- a coincidence
limited to that order. It is therefore the mean-square displacement along the
chosen direction, and it carries the diffusion coefficient, so that transport
is obtained from the ISF without the distribution ever being constructed.

The distribution \eqref{eq:rhoR} tends to $\theta$ and not to zero at large
separation, so \eqref{eq:cfdef} is a transform in the sense of distributions, its
uniform background and its decaying part transforming into the two parts of the
signal,
\begin{equation}
I(\K,t) \;=\; \underbrace{\theta\sum_{\R} e^{\,i\K\cdot\R}}_{A(\K)}
\;+\; \sum_{\R} \bigl[\rho_{\R}(t)-\theta\bigr]\, e^{\,i\K\cdot\R} .
\label{eq:combsplit}
\end{equation}
The first term is a Dirac comb at the reciprocal-lattice points, that is, the
Bragg peaks, the elastic
part, fixed by the coverage alone and carrying no dynamics. The second converges
absolutely and holds the whole of the time dependence. Away from the Bragg points,
the comb vanishes so that the second term alone is measured at every momentum
transfer an experiment samples.
The number of adsorbates is conserved and the signal tends to the
constant $F(\mathbf 0)$ as $\K\to\mathbf 0$. Thus,  \eqref{eq:cfdef} is a CF because $\rho_{\R}(t)\ge0$, whereas what is
measured is the transform of $\rho_{\R}(t)-\theta$, which takes either sign,
so that cumulants evaluated on the measured signal are those of a signed
weight and need not be positive. 

Two regimes of the ISF are known without any model of the layer. Let $\gamma$ be the friction the substrate exerts on the
adsorbate, so that $\gamma^{-1}$ is the time the substrate takes to deflect it.
At times shorter than $\gamma^{-1}$ (the ballistic regime), the adsorbates have not yet been
deflected and move freely. On expanding \eqref{eq:cfdef} in the time the first
order vanishes, forward and backward velocities being equally likely. In the second order only the contribution of an adsorbate with itself survives, the
velocities of two distinct adsorbates being uncorrelated in equilibrium; what
remains is the mean-square velocity along $\K$, so that the signal leaves the
amplitude \eqref{eq:Fdef} quadratically in the time and quadratically in the
momentum transfer,
\begin{equation}
I(\K,t) \;=\; F(\K)\;-\;\tfrac12\,\avg{v_{\parallel}^{2}}\,K^{2}t^{2}
\;+\;O(t^{4}),
\qquad t\ll\gamma^{-1} .
\label{eq:fsum}
\end{equation}
Which velocity enters depends on the regime~\cite{torres2026}. For a classical
adsorbate of mass $M$, equipartition fixes $\avg{v_{\parallel}^{2}}=k_BT/M$; for a
light one, in the quantum regime, the free motion is the spreading of the wave
packet and the velocity is $ v_s=\hbar/2M\sigma_0$, with $\sigma_0$ its
initial width, so that it carries no temperature. The friction enters neither
case, the motion being free, and no effective mass can be assigned to the
adsorbate here. The decrement in \eqref{eq:fsum} is independent of the structure
of the layer, which enters only through the amplitude from which the decay
starts; measured against that amplitude, \eqref{eq:fsum} is the beginning of a
Gaussian decay,
\begin{equation}
\frac{I(\K,t)}{F(\K)} \;\simeq\;
\exp\left[-\,\frac{\avg{v_{\parallel}^{2}}\,K^{2}t^{2}}{2\,F(\K)}\right],
\qquad t\ll\gamma^{-1},
\label{eq:ballistic}
\end{equation}
in the time, and in the momentum transfer as well whenever the amplitude is
flat, as it is for a single adsorbate, whose SSF is unity. 
Now, at times longer than $\gamma^{-1}$ (the diffusive regime), the adsorbate has been deflected many
times, its displacement is a sum of contributions the substrate has decorrelated,
the second cumulant of \eqref{eq:cumulants} grows linearly in the time and the
higher ones no faster. The decay
is exponential in the time, 
\begin{equation}
I(\K,t) \;\simeq\; B(\K)\, e^{-\alpha(\K)\,t},
\qquad t\gg\gamma^{-1} .
\label{eq:diffusive}
\end{equation}
Equation \eqref{eq:diffusive} is the expression to which a HeSE
measurement is fitted~\cite{Jardine2009}, together with the elastic baseline $A(\K)$ of
\eqref{eq:combsplit}: $\alpha(\K)$ is the dephasing rate it reports and $B(\K)$
the amplitude, which is not the $F(\K)$ from which \eqref{eq:fsum} starts.
Since by \eqref{eq:fsum} the true signal leaves $F(\K)$ with vanishing slope,
whereas a description in terms of jumps starts with a finite one, the origin
of time employed hereafter is the one at which the jump regime begins, and
short times are to be understood as short compared with the mean interval
between jumps. Two losses of the amplitude are important and they are of different kinds.
Part of the decay occurs before the jump regime begins and is not recovered by
any description that starts there; and within that description the amplitude
returned by fitting \eqref{eq:diffusive} is still not $F(\K)$, the fit being made
over the range in which the signal decays, not at its origin. What is
exact is $I(\K,0)=F(\K)$; the fitted $B(\K)$ lies below it, and the second of these losses is one of the quantities the memory function
determines. Between the two limits, the lateral interaction becomes manifest.

The elastic baseline is a Dirac comb after \eqref{eq:combsplit}  and so it vanishes at every momentum away from a reciprocal-lattice point, which is why the diffusive amplitude and the
SSF coincide there. The non-zero baseline a real fit returns is extrinsic,
coming from defects, immobile adsorbates and substrate background. The closed
form reached in Sec.~\ref{sec:dyn} has precisely the shape of \eqref{eq:diffusive},
with a rate fixed by the equilibrium layer alone and with the fitted amplitude
taken to be the diffusive part itself, $B(\K)=F(\K)$. What stands between it and
the exact decay is the equation of motion (memory equation) obeyed by \eqref{eq:cfdef}.

\subsection{The memory equation of the intermediate scattering function}
\label{sec:eom}

In the diffusive regime, the distribution $\rho_{\R}$ evolves by the jumps.  Whether an adsorbate can move to a
given neighbouring site, and how fast, depends on which of the surrounding sites
are occupied, and those occupations are not carried by $\rho_{\R}$: the
separation is one variable of the layer among many, and the rest were removed in
arriving at \eqref{eq:rhoR}. Eliminating them from the equation of motion obeyed
by $\hat\rho_{\rm red}$, by the projection of Nakajima and
Zwanzig~\cite{nakajima,zwanzig60}, leaves a master equation for $\rho_{\R}$ alone
in which the transfer between two lattice points is not instantaneous,
\begin{equation}
\frac{\partial \rho_{\R}(t)}{\partial t}
\;=\; \sum_{\R'}\int_0^t\! dt'\;\Bigl[\,
\mathcal K_{\R-\R'}(t-t')\,\rho_{\R'}(t')
\;-\;\mathcal K_{\R'-\R}(t-t')\,\rho_{\R}(t')\,\Bigr] ,
\label{eq:gme}
\end{equation}
with $\mathcal K_{\R-\R'}(\tau)$ the transfer kernel, the rate at which a layer
that held the separation $\R'$ a time $\tau$ earlier presents the separation
$\R$ now. Its two structural features come from properties already established: conservation of particles forces  gain and loss processes to enter with the
same kernel and opposite signs, since summing \eqref{eq:gme} over $\R$ must
vanish term by term, which is \eqref{eq:norm}. Spatial uniformity makes the
kernel depend on the change of separation $\mathbf L=\R-\R'$ alone. Equation
\eqref{eq:gme} is exact on two conditions that hold here: (i) that the projection is applied to a
linear equation of motion, which is what the equation for
$\hat\rho_{\rm red}$ is and (ii) that the layer starts in equilibrium, so that the
inhomogeneous term carrying the initial correlations between the separation and
the eliminated variables vanishes identically. A transfer made now depends on transfers made earlier, those earlier
moves having left the occupations around the adsorbate in a state that has not
yet relaxed (non-Markovian regime). In the Markovian regime, the kernel carries no delay,
$\mathcal K_{\mathbf L}(\tau)=\Gamma_{\mathbf L}\,\delta(\tau)$ with
$\Gamma_{\mathbf L}=\Gamma_{-\mathbf L}$ a set of rates which are constants, the convolution
collapses and \eqref{eq:gme} reduces to the gain and loss balance
\begin{equation}
\frac{\partial \rho_{\R}(t)}{\partial t}
\;=\; \sum_{\mathbf L}\Gamma_{\mathbf L}
\bigl[\,\rho_{\R+\mathbf L}(t)-\rho_{\R}(t)\,\bigr] ,
\label{eq:pauli}
\end{equation}
that is, the Pauli master equation for the diagonal elements is valid whether  the
jumps are thermally activated or proceed by tunnelling, the two entering through
the rates alone~\cite{torres2026}. The sum runs over the jump vectors of
\eqref{eq:jumps} and $\Gamma=\sum_{\mathbf L}\Gamma_{\mathbf L}$ is the total
rate at which one adsorbate leaves its site. Equation \eqref{eq:pauli} is exact
at vanishing coverage, where every neighbouring site is empty and the rates are
indeed constants of the layer; the retardation in \eqref{eq:gme} is what the
lateral interaction adds.

A lattice Fourier transform now carries \eqref{eq:gme} to the momentum
transfer space: multiplying by $e^{\,i\K\cdot\R}$ and summing over $\R$ --- the
same transform defined in \eqref{eq:cfdef} --- removes the
coupling between separations. In the gain term, the substitution
$\R=\R'+\mathbf L$ factorizes the double sum into a phase carried by $\mathbf L$
and the transform \eqref{eq:cfdef} carried by $\R'$; in the loss term, the sum
over $\R'$ produces the same kernel at zero momentum transfer. One scalar
equation is left at each momentum transfer,
\begin{equation}
\frac{\partial I(\K,t)}{\partial t}
\;=\; -\int_0^t\! dt'\; \mathcal W(\K,t-t')\, I(\K,t') ,
\qquad
\mathcal W(\K,\tau) \;=\; \sum_{\mathbf L}\bigl[\,1-e^{\,i\K\cdot\mathbf L}\,\bigr]
\,\mathcal K_{\mathbf L}(\tau) ,
\label{eq:volterra}
\end{equation}
The factor
$1-e^{\,i\K\cdot\mathbf L}$ is the difference between the loss and gain phases
and vanishes at $\K=\mathbf 0$ term by term, so that $\mathcal W(\mathbf 0,\tau)=0$
and the weight of the signal stays constant: the conservation law used in \eqref{eq:norm} is returned by the dynamics,
not imposed on it. At $t=0$,
the integral is empty so that $I(\K,0)=F(\K)$ is untouched and the amplitude of
the signal is fixed by the equilibrium layer and not by its evolution.

One further CF is required before \eqref{eq:volterra}
can be interpreted. Let $w_{\mathbf L}$ be the probability that a single jump is the one with
jump vector $\mathbf L$ of \eqref{eq:jumps}, so that $w_{\mathbf L}\ge0$ and
$\sum_{\mathbf L}w_{\mathbf L}=1$, and let
\begin{equation}
f_{\K} \;=\; \sum_{\mathbf L} w_{\mathbf L}\, e^{\,i\K\cdot\mathbf L}
\label{eq:fK}
\end{equation}
be the CF of that single-jump distribution, evaluated at the
momentum transfer of the measurement; \eqref{eq:fK} is to one jump what
\eqref{eq:cfnorm} is to the accumulated displacement. Forward and backward jumps
are equally likely in a uniform layer, $w_{\mathbf L}=w_{-\mathbf L}$, so that
$f_{\K}$ is real and even in $\K$ and every rate built from it below is real. In
the Markovian regime, the jump probabilities are
$w_{\mathbf L}=\Gamma_{\mathbf L}/\Gamma$, the kernel of \eqref{eq:volterra} is
$\mathcal W(\K,\tau)=\Gamma[1-f_{\K}]\delta(\tau)$, the equation is local, and its
solution is a single exponential,
\begin{equation}
I(\K,t) \;=\; F(\K)\,e^{-\Gamma\,[1-f_{\K}]\,t} ,
\label{eq:ce}
\end{equation}
the model of Chudley and Elliott~\cite{Chudley1961}, which as a CF is the compound Poisson form of independent increments whose number in a
time $t$ is Poisson distributed with mean $\Gamma t$~\cite{Montroll1965}. For
NN jumps along one lattice direction, \eqref{eq:fK} gives
$f_{\K}=\cos(K a\cos\tilde\beta)$ and longer jumps add higher harmonics.

The two contributions to $\mathcal W$ are not on the same footing, the jumps
being instantaneous on the scale of the interval that separates them. Part of the
transfer therefore occurs with no delay elapsed and the rest carries the
dependence on transfers made earlier, so that at each momentum transfer
\begin{equation}
\mathcal W(\K,\tau) \;=\; \Omega(\K)\,\delta(\tau)\;-\;\Mc(\K,\tau) ,
\label{eq:split}
\end{equation}
the sign of the second term being chosen so that a positive memory function slows
the decay. With \eqref{eq:split}, the Volterra equation becomes
\begin{equation}
\frac{\partial I(\K,t)}{\partial t}
\;=\; -\,\Omega(\K)\, I(\K,t) \;+\; \int_0^t\! dt'\; \Mc(\K,t-t')\, I(\K,t') ,
\label{eq:eom}
\end{equation}
which is the form Mori gave to the equation of motion of a single dynamical
variable~\cite{Mori1965,Zwanzig2001}. The two terms now play distinct
dynamical roles. $\Omega(\K)$ is the local part of the transfer, the fraction that acts
instantaneously (during the first instants of the jump
regime) before any retardation has had time to accumulate. It is the whole of
the dynamics, the convolution there being still empty. Differentiating at that
origin --- which since Sec.~\ref{sec:cf} is the instant at which the jump
regime begins, and not the laboratory origin, where \eqref{eq:fsum} makes the
slope vanish --- therefore returns the local rate alone,
\begin{equation}
	\Omega(\K) \;=\; -\,\frac{\partial \ln I(\K,t)}{\partial t}
	\bigg|_{t=0} ,
	\label{eq:Omega}
\end{equation}
$\Omega(\K)$ being thus the rate at which the decay of the density wave sets
out; the kernel acts only at later times, and with the sign convention of
\eqref{eq:split} a positive memory can only slow the decay thereafter. The local term initiates the relaxation,
the retarded term moderates it. 
Nothing has been approximated between
\eqref{eq:rhoR} and \eqref{eq:eom}. Discarding $\Mc$ makes \eqref{eq:eom}
Markovian at all times and returns a single exponential of rate $\Omega(\K)$ with
the exact amplitude \eqref{eq:Fdef}; the memory function is retained here.

\subsection{The sum rule}
\label{sec:coeff}

Equation \eqref{eq:eom} is exact but does not determine its own coefficients:
the transform of \eqref{eq:gme} does not provide $\Omega(\K)$, and
the form of \eqref{eq:eom} does not decide the sign of $\Mc$. Obtaining them
requires the layer itself to be followed, and not the separation alone.
The key concept now is the configuration
which represents the set of occupation numbers introduced in Sec.~\ref{sec:cf} and the
moves being exchanges between an occupied site and an empty neighbouring
one~\cite{Kawasaki1966}, which conserve the number of adsorbates. What the steps
below use is not the form of the lateral interaction but only that the rates
carry it through the equilibrium measure. 
One can then proceed as follows: 

The first step identifies the combination of the occupations  and it is fixed by the phase $e^{\,i\K\cdot\R}$ that the
scattered wave attaches to each site. Measuring
each occupation from its mean through the occupation fluctuation
\begin{equation}
\delta n_{\R} \;\equiv\; n_{\R}-\theta ,
\label{eq:deltan}
\end{equation}
that phase weights the layer with the density wave of wavevector $\K$,
\begin{equation}
\delta n_{\K} \;=\; \sum_{\R}\delta n_{\R}\,e^{\,i\K\cdot\R} ,
\label{eq:densitywave}
\end{equation}
whose mean-square value per site is the density--density structure factor
$S(\K)=\avg{|\delta n_{\K}|^{2}}/N$, with $N$ the number of sites. In the same
variables, the conditioning of Sec.~\ref{sec:cf} is an average of the occupation of the
site $\R$ at time $t$ over the layers in which the origin was occupied at time
zero,
\begin{equation}
\rho_{\R}(t) \;=\; \frac{\avg{\,n_{\mathbf 0}(0)\,n_{\R}(t)\,}}{\theta} ,
\label{eq:rhoocc}
\end{equation}
the denominator being the probability that the origin is occupied at all, and
returning $\rho_{\mathbf 0}(0)=1$ and $\rho_{\R}(0)\to\theta$ as required.
Subtracting the uniform background as in \eqref{eq:combsplit} replaces the
occupations by their fluctuations, and the origin may then be averaged over the
$N$ sites, every one of which is equivalent in a uniform layer,
\begin{equation}
I(\K,t)\;=\;\sum_{\R}\bigl[\rho_{\R}(t)-\theta\bigr]e^{\,i\K\cdot\R}
\;=\;\frac{1}{\theta}\sum_{\R}\avg{\,\delta n_{\mathbf 0}(0)\,\delta n_{\R}(t)\,}
e^{\,i\K\cdot\R}
\;=\;\frac{\avg{\,\delta n_{-\K}(0)\,\delta n_{\K}(t)\,}}{N\theta} .
\label{eq:Iautocorr}
\end{equation}
What a coherent measurement records is therefore the equilibrium autocorrelation
of a single density wave, whose value at $t=0$ is fixed by the statics alone,
\begin{equation}
I(\K,0)\;=\;\frac{\avg{|\delta n_{\K}|^{2}}}{N\theta}\;=\;\frac{S(\K)}{\theta}
\;=\;F(\K) ,
\label{eq:FSrel}
\end{equation}
which is \eqref{eq:Fdef} written in the occupations, the two differing by the
factor $\theta$ because $F(\K)$ transforms the conditional pair distribution
$P_{\R}$ and $S(\K)$ the correlation of the fluctuations. Now, let $\Lc$ denote the generator of the
process,
so that a quantity carried a time $t$ forward is $e^{\Lc t}$ acting on it, and
\eqref{eq:gme} is what the same evolution becomes once every variable but the
separation has been eliminated. The rates obey detailed balance: in equilibrium,
the flux from one state of the layer to another is matched by the reverse flux
for every pair separately, so that the layer in equilibrium is statistically the same whether the evolution is run forwards or backwards. That is what reversibility
means here. It is a property of the substrate that drives the moves, not an assumption
about the lateral interaction, and its consequence is that $\Lc$ is
self-adjoint and non-positive in the equilibrium measure~\cite{liggett}.
Since the layer is carried forward by $\Lc$, \eqref{eq:Iautocorr} becomes
\begin{equation}
I(\K,t) \;=\; \frac{\avg{\,\delta n_{-\K}\;e^{\Lc t}\,\delta n_{\K}\,}}{N\theta},
\qquad I(\K,0)=F(\K),
\label{eq:regression}
\end{equation}
the average now running over the equilibrium measure alone.

The second step now extracts from \eqref{eq:regression} the two coefficients
left open by \eqref{eq:eom}, and the tool is an orthogonal projection. The observables of the layer form a vector space on which the
equilibrium measure defines a scalar product, $(A,B)\equiv\avg{A^{*}B}$, with
$(\delta n_{\K},A)=\avg{\delta n_{-\K}\,A}$; it is in this product that
reversibility makes $\Lc$ self-adjoint and non-positive, and in it the density
wave spans a single direction. The instantaneous change of the 
variable, $\Lc\,\delta n_{\K}$, is itself an observable of the layer, and like
any vector it admits a unique decomposition into a component along
$\delta n_{\K}$ and a remainder orthogonal to it,
\begin{equation}
\Lc\,\delta n_{\K} \;=\; -\,\Omega(\K)\,\delta n_{\K}\;+\;\phi_{\K},
\qquad
\avg{\delta n_{-\K}\,\phi_{\K}} = 0 ,
\qquad
\Omega(\K) \;=\; -\,\frac{\avg{\delta n_{-\K}\,\Lc\,\delta n_{\K}}}
{\avg{|\delta n_{\K}|^{2}}} ,
\label{eq:decomp}
\end{equation}
in which nothing has been chosen: the coefficient is the projection of
$\Lc\,\delta n_{\K}$ on the direction of the measurement, real because
$\Lc$ is self-adjoint and non-negative because $\Lc$ is non-positive, and the
remainder $\phi_{\K}$ is the part of the instantaneous dynamics that the
 variable cannot express. Differentiating \eqref{eq:regression} at
the origin of the jump description returns \eqref{eq:Omega}, so that the two
definitions of $\Omega(\K)$ agree; the second is an equilibrium
average. However, a rate extracted from a decay is not, and it is what makes the split \eqref{eq:split} a consequence, not an
assumption. Carried through the evolution, the decomposition \eqref{eq:decomp}
identifies the memory function of \eqref{eq:eom} as the autocorrelation of
that orthogonal remainder,
\begin{equation}
\Mc(\K,\tau) \;=\; \frac{\avg{\,\phi_{-\K}\;e^{\Lc_{\perp}\tau}\,\phi_{\K}\,}}
{\avg{|\delta n_{\K}|^{2}}} ,
\label{eq:memdef}
\end{equation}
with $\Lc_{\perp}$ the generator restricted to the space orthogonal to the
density wave; the operator identity behind \eqref{eq:memdef} is due to Mori, and its
derivation is standard~\cite{Mori1965,Zwanzig2001}. 
Two of its consequences are used throughout, and both rest on the self-adjointness and non-positivity that $\Lc_{\perp}$ inherits from $\Lc$ on
that subspace. The memory function is non-negative --- indeed itself a
superposition of decaying exponentials with non-negative weights, reducing at
zero delay to $\Mc(\K,0)=\avg{|\phi_{\K}|^{2}}/\avg{|\delta n_{\K}|^{2}}$ ---
so that a positive memory can only oppose the decay, which is what
\eqref{eq:split} was written to display. And its moments are equilibrium
averages of the layer, static quantities computable without following the
dynamics at all.

Finally, the third step evaluates the projection: reversibility turns the numerator
of $\Omega(\K)$ into a manifestly non-negative equilibrium average. 
A move carrying an adsorbate from $\R$ to $\R+\mathbf L_j$ removes a
unit at $\R$ and adds one at $\R+\mathbf L_j$, so that by
\eqref{eq:densitywave} the density wave changes by
$e^{\,i\K\cdot\R}\bigl(e^{\,i\K\cdot\mathbf L_j}-1\bigr)$. Its modulus
squared, $2[1-\cos(\K\cdot\mathbf L_j)]$, is a pure number: the site phase has
unit modulus and drops out, so the change is the same in every configuration
and for every site that moves. Thus, only the rate is left to average,
\begin{equation}
-\,\avg{\,\delta n_{-\K}\,\Lc\,\delta n_{\K}\,}
\;=\;\sum_{\R,\,j}\bigl\langle\, w_{\R\to\R+\mathbf L_j}\;
n_{\R}\,\bigl(1-n_{\R+\mathbf L_j}\bigr)\,\bigr\rangle\,
\bigl[\,1-\cos(\K\cdot\mathbf L_j)\,\bigr] ,
\label{eq:dirichlet}
\end{equation}
the occupancies appearing because a move happens only if the departure site is
occupied and the target empty. Every term of that average is the same number,
by translational invariance and by the equivalence of the $z$ neighbour
directions of a uniform layer. If this is represented by $J$ then 
\begin{equation}
J \;\equiv\; \bigl\langle\, w_{\R\to\R+\mathbf L_j}\;
n_{\R}\,\bigl(1-n_{\R+\mathbf L_j}\bigr)\,\bigr\rangle .
\label{eq:current}
\end{equation}
It is the mean number of exchanges a given bond performs per unit time in one
direction: the rate at which an adsorbate sitting at one end succeeds in moving
to the other. It is not a net current --- detailed balance makes the two
directions of a bond carry equal flux, so their difference vanishes in
equilibrium --- but one way flux, which is what a rate must be.
Everything the interaction does to the dynamics sits inside it: the rate $w$
depends on the surrounding occupations, and whether the move is possible at all
depends on the two sites of the bond, and both are averaged together over the
same equilibrium measure. Summing \eqref{eq:dirichlet} over the $N$ sites and
the $z$ directions, with $\sum_j[1-\cos(\K\cdot\mathbf L_j)]=z[1-f_{\K}]$ and
$\avg{|\delta n_{\K}|^{2}}=N\theta F(\K)$ by \eqref{eq:FSrel} (the sum rule)
\begin{equation}
\Omega(\K)\,F(\K) \;=\; \frac{zJ}{\theta}\,\bigl[\,1-f_{\K}\,\bigr]
\;\equiv\;\Gamma_{\rm eff}\,\bigl[\,1-f_{\K}\,\bigr] ,
\label{eq:sumrule}
\end{equation}
where $\Gamma_{\rm eff}=zJ/\theta$ is the rate at which one adsorbate leaves its
site: $J$ counts exchanges per bond, $z$ directions are open to it, and
dividing by $\theta$ turns a rate per site into a rate per adsorbate. Should jumps of more than
one length be present, each length carries its own current and the weights of
\eqref{eq:fK} are the ratios of those currents.
Equation \eqref{eq:sumrule} carries no structure. Its momentum dependence is the
single-jump geometry $1-f_{\K}$ alone, and the interaction enters it only through
the scalar $\Gamma_{\rm eff}$. All the structure of the layer resides in $F(\K)$, and therefore in the
denominator of the rate,
$\Omega(\K)=\Gamma_{\rm eff}[1-f_{\K}]/F(\K)$, so that relaxation slows wherever the adlayer has built up structure and quickens
where structure is suppressed. This is the narrowing described by de
Gennes~\cite{deGennes1959} for the coherent line width of a dense fluid, and on the lattice it is a theorem, not an analogy, since \eqref{eq:sumrule}
fixes the product $\Omega(\K)F(\K)$ without ever using the form of $F(\K)$.

The passage from \eqref{eq:decomp} to \eqref{eq:sumrule} uses neither the Ising
form of the interaction, nor its range, nor any closed form for the structure
factor. It uses two properties only: that the moves are exchanges between a fixed
set of neighbour vectors, and that the rates obey detailed balance with respect
to whatever Hamiltonian governs the layer. Then \eqref{eq:current} is the
exact equilibrium exchange flux, so that $\Gamma_{\rm eff}$ is defined by \eqref{eq:sumrule}, not
approximated, and \eqref{eq:sumrule} is exact.
Moreover,  \eqref{eq:sumrule} defines $\Gamma_{\rm eff}$ but does not evaluate it.
Evaluating \eqref{eq:current} requires the equilibrium statistics of a particular
layer, which belongs to the application of the theory, not to its construction; supplied with it, \eqref{eq:sumrule} becomes an elementary function
of momentum transfer, coverage and temperature, and that function is the
dephasing rate. 
What the present subsection
adds is its status. It is the exact local coefficient of an exact equation of motion, and the factor
$1/F(\K)$ is the theorem \eqref{eq:sumrule}, not a closure imposed on the
kinetics. Whatever approximation the evaluation of
$\Gamma_{\rm eff}$ carries affects a scalar and not the momentum dependence.

The accumulated effect of the memory on the measured signal is itself
measurable. The area under the normalized decay defines the correlation time
$\tau_c(\K)=\int_0^{\infty}\!dt\,I(\K,t)/F(\K)$ and, in the Laplace domain,
the saturated memory is the difference between the rate at which the decay
starts and the rate the area assigns to it,
$\hat\Mc(\K,0)=\Omega(\K)-1/\tau_c(\K)$; the memory being non-negative, that
difference is too, and $\Omega(\K)\,\tau_c(\K)\ge1$, with equality if and only
if the memory vanishes identically.
The same positivity orders the rates. Dividing \eqref{eq:eom} by $I(\K,t)$ gives
the instantaneous decay rate as
\begin{equation}
-\,\frac{\partial \ln I(\K,t)}{\partial t}
\;=\; \Omega(\K)\;-\;\frac{1}{I(\K,t)}\int_0^t\! dt'\;\Mc(\K,t-t')\,I(\K,t') ,
\label{eq:instrate}
\end{equation}
in which the subtracted term is non-negative at every time, so that the signal
never decays faster than $\Omega(\K)$. An exponential fitted over any range of
times therefore returns $\alpha(\K)<\Omega(\K)$, and how far below depends on the
range fitted. Beyond the range of $\Mc$, the convolution has saturated and
\eqref{eq:eom} is Markovian once more, but with the coefficient reduced from
$\Omega(\K)$ to $1/\tau_c(\K)$; within that range the decay is not exponential at
all.
Discarding $\Mc$ altogether returns
\begin{equation}
I_1(\K,t) \;=\; F(\K)\,e^{-\Omega(\K)\,t} ,
\label{eq:level1}
\end{equation}
the subscript marking it as the first truncation of \eqref{eq:eom}, with both
of its coefficients derived rather than posited, the amplitude by
\eqref{eq:regression} and the rate by \eqref{eq:sumrule}. It is exact in amplitude and in slope at the origin
of the jump description, and nowhere else; and since a positive memory can only
oppose the decay, it lies below the exact signal at every later time.

That construction has been used once already. At the ballistic end the first time
derivative that does not vanish is the second, and \eqref{eq:fsum} shows it to be insensitive to the structure of the layer, so that the Gaussian \eqref{eq:ballistic}
reproducing it is fixed by the amplitude and by that derivative alone. In the
regime, the first derivative that does not vanish is the first, and
\eqref{eq:sumrule} shows it to be insensitive to the structure as well, so that the
exponential \eqref{eq:level1} is fixed by the same two data. The order of the
surviving derivative is what distinguishes the two ends; in both, the rate is
that derivative divided by the amplitude, which is why the SSF
appears divided in \eqref{eq:ballistic} and in \eqref{eq:sumrule} alike. What
\eqref{eq:level1} does not reproduce is the order beyond, and the size of that
failure at the origin is $\Mc(\K,0)$.

\subsection{The intermediate scattering function as a superposition of
relaxation modes}
\label{sec:modes}

Reversibility, which made the local coefficient of \eqref{eq:eom} a
non-negative equilibrium average, also fixes the shape of the decay. Being self-adjoint and
non-positive in the equilibrium measure, $\Lc$ possesses a complete set of
eigenvectors with real, non-positive eigenvalues $-\lambda_k$. Expanding the
density wave in that set turns the regression \eqref{eq:regression} into a
superposition of exponentials,
\begin{equation}
I(\K,t) \;=\; \sum_k a_k(\K)\, e^{-\lambda_k t},
\qquad \lambda_k\ \ge\ 0,
\qquad a_k(\K)\ \ge\ 0 ,
\label{eq:modes}
\end{equation}
each mode decaying at its own rate and carrying a weight that cannot be negative,
being the squared overlap of the density wave of wavevector $\K$ with that mode.
Setting $t=0$,
\begin{equation}
\sum_k a_k(\K) \;=\; F(\K),
\label{eq:weights}
\end{equation}
so that the SSF fixes how the signal is distributed among the
modes and
$a_k(\K)/F(\K)$ is the fraction carried by one of them. Equation \eqref{eq:modes}
holds at times $t \gg \gamma^{-1}$, its origin of time being the one fixed in
Sec.~\ref{sec:cf}. The derivatives taken at that origin below are those of the
jump description and not of the measured signal at the laboratory origin, where
\eqref{eq:fsum} makes the slope vanish. On an infinite lattice, the modes are
dense and the sum becomes an integral over rates, which changes nothing below.

The expansion replaces a factorization that is available only in the dilute
limit. A CF handles independent increments by multiplying them, which
is what produced the compound Poisson form \eqref{eq:ce} and its single
exponential; when the increments are correlated that product fails, and
\eqref{eq:modes} takes its place. It also supplies an expansion in the time where
there was none. In the momentum transfer, the CF has always
generated the moments of the displacement, and \eqref{eq:modes} lets it do the
same in the time, so that the memory function acquires a structure of its own
instead of remaining a kernel with none. And it converts a constraint that could
not be used into one that can: positive definiteness in $\K$, which is what makes
\eqref{eq:cfnorm} a CF, constrains the decay in the time only weakly, whereas non-negative weights
in \eqref{eq:modes} state that the decay is completely monotone, and the
inequalities of the remainder of this work rest on that and on nothing else.

In this language, the ballistic regime of Sec.~\ref{sec:cf} lies outside the
expansion altogether, a Gaussian in the time not being a superposition of
decaying exponentials with non-negative weights. Within the diffusive regime,
where \eqref{eq:modes} holds, the question is instead how many modes remain
active. At times long compared with every mode, only the slowest survives,
the sum is left with a single term, and the decay is the single exponential
\eqref{eq:diffusive} to which data are fitted. Earlier than that, several
modes contribute at once, and that is the whole of what the memory function
describes:
$\Mc$ records the presence of more than one active mode.
The ISF thereby admits a double description. In the momentum transfer, it is the CF of the displacement, Eq.~\eqref{eq:cfnorm}; in the time,
it is a superposition of relaxation modes, Eq.~\eqref{eq:modes}. The first tells us
how the adsorbate is spread over positions, the second how the signal is spread
over rates, and in each case the moments of the spread are derivatives of the ISF
at the origin of the corresponding variable.
One consequence is immediate. The instantaneous decay rate of the signal,
\begin{equation}
-\,\frac{\partial \ln I(\K,t)}{\partial t}
\;=\; \frac{\sum_k \lambda_k\, a_k(\K)\, e^{-\lambda_k t}}
           {\sum_k a_k(\K)\, e^{-\lambda_k t}} ,
\label{eq:inst}
\end{equation}
is the average of the mode rates weighted by what each mode still contributes at
that time. As time passes, the fast modes decay first, the average shifts
towards the slow ones, and the instantaneous rate can only fall. At the origin,
every exponential is unity and it reduces to
\begin{equation}
\frac{\sum_k \lambda_k\, a_k(\K)}{\sum_k a_k(\K)}
\;=\; \Omega(\K) ,
\label{eq:meanrate}
\end{equation}
which identifies the local coefficient of
\eqref{eq:eom} as the mean of the mode rates weighted by the part of the signal
each mode carries, a property of the equilibrium layer, not of the point at which the
description is started. Equations \eqref{eq:inst} and
\eqref{eq:meanrate} return in arithmetic form what \eqref{eq:memdef} gave
microscopically, that the memory function is non-negative and that an exponential
fitted over any range of times returns a rate below $\Omega(\K)$; what the mode
form adds is the monotonicity, which the equation of motion alone does not
supply.

Differentiating \eqref{eq:modes} at the origin gives the moments of the mode
rates, each weighted by the part of the signal its mode carries,
\begin{equation}
\omega_n(\K) \;\equiv\; (-1)^n
\frac{\partial^{\,n} I(\K,t)}{\partial t^{\,n}}\bigg|_{t=0}
\;=\; \sum_k a_k(\K)\,\lambda_k^{\,n} ,
\label{eq:moments}
\end{equation}
with $\omega_0(\K)=F(\K)$ by \eqref{eq:weights}. These are moments of a
distribution of rates, and are not the moments of the displacement of
\eqref{eq:cumulants}, which are derivatives in $K$ and not in $t$.
The first two are fixed by the equation of motion. At $t=0$, the convolution in
\eqref{eq:eom} is empty, so that $\omega_1(\K)=\Omega(\K)F(\K)$, which is
\eqref{eq:meanrate} once more. Differentiating \eqref{eq:eom} a second time and
setting $t=0$, where the convolution now contributes $\Mc(\K,0)F(\K)$,
\begin{equation}
\omega_2(\K) \;=\; \bigl[\,\Omega^{2}(\K)+\Mc(\K,0)\,\bigr]\,F(\K) ,
\label{eq:om2}
\end{equation}
so that the memory function starts at the spread of the mode rates about their
mean,
\begin{equation}
\Mc(\K,0) \;=\; \Delta(\K)\,\Omega^{2}(\K),
\qquad
\Delta(\K) \;=\; \frac{\omega_0(\K)\,\omega_2(\K)}{\omega_1^{2}(\K)} - 1
\;\ge\; 0 .
\label{eq:Delta}
\end{equation}
Here $\Delta(\K)$ is the variance of the mode rates divided by the square of
their mean --- their squared relative width --- non-negative because the
weights in \eqref{eq:modes} are, and vanishing when one rate carries the whole
signal and only then: a layer relaxing through a single
mode has no memory and decays as one exponential, whereas when several modes
share the signal they decay at different rates, the mixture is not an
exponential, and the memory function is what accounts for the difference.

The truncation \eqref{eq:level1} is thereby controlled from both sides. Its sign
follows from the equation of motion alone: the subtracted term in
\eqref{eq:instrate} is non-negative at every time, so that integrating that
equation from the origin of the jump description gives
\begin{equation}
I(\K,t) \;\ge\; F(\K)\,e^{-\Omega(\K)\,t}
\qquad\text{at every } \K \text{ and every } t ,
\label{eq:bound}
\end{equation}
and the closed form is a rigorous lower bound on the ISF, not an approximation of uncontrolled
sign. Its accuracy follows from the mode expansion.
Equation \eqref{eq:level1} is the single-mode representation reproducing
$\omega_0$ and $\omega_1$, that is, the whole distribution of rates replaced by
its mean, while the exact signal is the average of $e^{-\lambda t}$ over that
distribution, which convexity places above the exponential of the mean. Comparing
the two,
\begin{equation}
\frac{I(\K,t)}{I_1(\K,t)} \;=\; \avg{\,e^{-[\lambda-\Omega(\K)]t}\,}
\;=\; 1+\tfrac12\,\Delta(\K)\,\Omega^{2}(\K)\,t^{2}+O(t^{3}) ,
\label{eq:excess}
\end{equation}
so that the relative excess of the exact signal over the closed form is one half
of $\Delta$ times the square of the scaled time $\Omega t$. This has two consequences. Wherever $\Delta(\K)$ vanishes, the Markovian truncation is exact, at
every momentum transfer and at every time, and not merely to leading order in the
interaction. And where it does not vanish, the closed form remains accurate over
the range in which the signal decays provided $\Delta$ is small: at
$\Omega(\K)t=1$, an excess of ten per cent corresponds, to leading order, to
$\Delta(\K)\simeq0.2$. The accuracy of the closed form is therefore carried by a
single static number at each momentum transfer.
The two moments that the truncation uses have a common microscopic origin.
Self-adjointness turns $\omega_2$, like $\omega_1$, into a manifestly
positive equilibrium average,
\begin{equation}
\omega_2(\K) \;=\; \frac{\avg{\,\bigl|\Lc\,\delta n_{\K}\bigr|^{2}\,}}{N\theta} ,
\label{eq:om2static}
\end{equation}
the lattice transform of the equilibrium correlation between the exchanges on two bonds.
The parallel with \eqref{eq:sumrule} is exact and it
fixes what each moment knows.
Both are static
equilibrium averages, so that $\Delta(\K)$, and with it the accuracy of the
truncation, is a property of the equilibrium layer and requires no dynamics
to be computed.

The condition under which the closed form ceases to be an approximation comes
from \eqref{eq:Delta}. Since $\Delta(\K)$ vanishes only when a single rate
carries the whole signal, and since the weights are squared overlaps,
$\Delta(\K)=0$ if and only if $\delta n_{\K}$ is itself an eigenvector of $\Lc$.
In other words, if and only if the generator maps the single-adsorbate density onto
itself. That happens whenever the rate of a move does not depend on the
occupations of the surrounding sites. For exchanges at a common rate
$\Gamma_0/z$ between an occupied site and an empty neighbour, the terms quadratic
in the occupations cancel between the gain and the loss,
\begin{equation}
\Lc\,n_{\R} \;=\; \frac{\Gamma_0}{z}\sum_j\bigl(n_{\R+\mathbf L_j}-n_{\R}\bigr) ,
\qquad
\Lc\,\delta n_{\K} \;=\; -\,\Gamma_0\bigl[1-f_{\K}\bigr]\,\delta n_{\K} ,
\label{eq:sep}
\end{equation}
so that the density wave is an exact eigenvector at every momentum transfer, one
mode carries the whole signal, $\Delta(\K)\equiv0$, and the closed form is exact, not merely accurate. Any lateral interaction that makes the escape rate depend
on the surroundings destroys that linearity and opens $\Delta(\K)>0$.

Two measurable manifestations of the memory are in order. First, in the frequency domain,
the dynamic structure factor is the transform of \eqref{eq:modes}, a
superposition of Lorentzians weighted by the same distribution of rates, whose
height at zero frequency is $F(\K)\tau_c(\K)/\pi$ against $F(\K)/\pi\Omega(\K)$
for the closed form: the ratio of the two is $\Omega(\K)\tau_c(\K)$, so that an
energy-resolved measurement returns that quantity from the peak height alone. Second, in the time domain, the instantaneous rate \eqref{eq:inst} falls
monotonically, so
that an exponential fitted over a wider window necessarily returns a smaller rate
than one fitted over a narrower window at the same momentum transfer. The sign of that drift is a theorem and its size is governed by
$\Delta(\K)$.

At long wavelength, the same distinction fixes the diffusion coefficient. As
$\K\to\mathbf 0$, the local rate \eqref{eq:sumrule} vanishes as $K^{2}$, since
$1-f_{\K}\to\tfrac12\avg{\ell^{2}}K^{2}$ with $\avg{\ell^{2}}$ the mean-square
projected jump length, whereas the memory function relaxes on the timescale of
the modes orthogonal to the density, which stays finite. Provided the Laplace transform 
$\hat\Mc(\K,s)$ is regular at $s\to0$ in that limit, which is the hydrodynamic
assumption and the only one made here, it may be replaced by
$\hat\Mc(\K,0)$: 
the equation becomes Markovian once more
and the decay proceeds at $1/\tau_c(\K)$ instead of at $\Omega(\K)$. With
$\alpha\to\tfrac12 D_cK^{2}$ and $\Gamma_{\rm eff}$ from \eqref{eq:sumrule},
\begin{equation}
D_c \;=\; \frac{\avg{\ell^{2}}\,\Gamma_{\rm eff}}{F(\mathbf 0)\,R(\mathbf 0)},
\qquad
R(\K)\;\equiv\;\Omega(\K)\,\tau_c(\K)\;\ge\;1 ,
\label{eq:Dhydro}
\end{equation}
the numerator being the jump diffusion coefficient of a layer hopping at the mean
rate. The factor $1/F(\mathbf 0)$ is the thermodynamic one, the inverse
compressibility, and gives the Darken relation; the factor $1/R(\mathbf 0)$ is
the ratio between the Onsager coefficient of the mean-rate description and the
true collective one, and it is exactly the correlation between successive moves
that a mean-rate treatment sets to unity. By \eqref{eq:sep} it is unity for the
non-interacting layer at every coverage, which is why the closed form returns the
symmetric exclusion process exactly; away from that limit it is a number the
layer supplies.

\subsection{Linear response and line shape}
\label{sec:lineshape}

The ISF then takes the single-exponential form \eqref{eq:diffusive},
$I(\mathbf{K},t)=A(\mathbf{K})+B(\mathbf{K})e^{-\alpha t}$, with amplitude
$B(\mathbf{K})=F(\mathbf{K})$ away from the Bragg points and rate
$\alpha(\mathbf{K})=\Omega(\mathbf{K})$ of \eqref{eq:sumrule} --- that is, the
truncation \eqref{eq:level1}. The whole hierarchy of linear-response
functions built on the ISF \cite{TorresMiyares2026c} follows from it by
linear, invertible operations, each member an experimentally distinct route
to the same relaxation, and every function below therefore inherits the
status of that truncation: exact in amplitude, exact in initial slope by the
sum rule \eqref{eq:sumrule}, and approximate only in shape, by the margin
bounded in Sec.~\ref{sec:modes}.

The primary spectral observable is the dynamic structure factor $S(\mathbf{K},\omega)$, the time Fourier transform of the ISF. The elastic baseline contributes a sharp line at zero frequency and the decaying part a Lorentzian centred at $\omega=0$, of full width at half maximum $2\alpha(\mathbf{K})$ and area $B(\mathbf{K})$,
\begin{equation}
	S(\mathbf{K},\omega) \;=\; A(\mathbf{K})\,\delta(\omega)
	\;+\; \frac{B(\mathbf{K})}{\pi}\;\frac{\alpha(\mathbf{K};\theta,T)}{\omega^{2}+\alpha^{2}(\mathbf{K};\theta,T)} .
	\label{eq:dsf}
\end{equation}
Integrated over frequency, it returns $I(\mathbf{K},0)=A(\mathbf{K})+B(\mathbf{K})$, which away from the Bragg points is $F(\mathbf{K})$: the area and the width of the measured peak are the equilibrium structure and the relaxation rate.


The time-domain response functions are written in the canonical (Kubo)
normalization, where $\beta\hbar$ carries the units \cite{TorresMiyares2026c}. The relaxation function is the regression of a density fluctuation prepared at $t=0$ --- the decaying part of the ISF itself, $R_{\rm rel}(\mathbf{K},t)=\beta\hbar\,F(\mathbf{K})\,e^{-\alpha(\mathbf{K})t}$ --- and minus its time derivative is the after-effect function $\phi(\mathbf{K},t)=\beta\hbar\,\alpha(\mathbf{K})F(\mathbf{K})\,e^{-\alpha(\mathbf{K})t}$. The generalized dynamic susceptibility is the one-sided transform of $\phi$, a single Debye form
\begin{equation}
	\chi(\mathbf{K},\omega) \;=\; \beta\hbar\,F(\mathbf{K})\,\frac{\alpha(\mathbf{K})}{\alpha(\mathbf{K})-i\omega},
	\label{eq:chi}
\end{equation}
whose real and imaginary parts,
\begin{equation}
	\chi'(\mathbf{K},\omega)=\beta\hbar\,F(\mathbf{K})\,\frac{\alpha^{2}(\mathbf{K})}{\alpha^{2}(\mathbf{K})+\omega^{2}},
	\qquad
	\chi''(\mathbf{K},\omega)=\beta\hbar\,F(\mathbf{K})\,\frac{\alpha(\mathbf{K})\,\omega}{\alpha^{2}(\mathbf{K})+\omega^{2}},
	\label{eq:chiparts}
\end{equation}
are the reactive and dissipative responses. The loss part $\chi''$ is a Debye absorption peak centred at $\omega=\alpha(\mathbf{K})$, so the relaxation rate is the frequency of maximum dissipation; the reactive part falls from its static value $\chi(\mathbf{K},0)=\beta\hbar F(\mathbf{K})$, the isothermal density response, to zero at high frequency, with Kramers--Kronig connecting the two. All three carry the same single rate $\alpha(\mathbf{K})$: writing $S_{\rm d}(\mathbf{K},\omega)$ for the decaying part of \eqref{eq:dsf}, the reactive response has its Lorentzian shape, $\chi'=\pi\beta\hbar\,\alpha\,S_{\rm d}$, while the loss part differs from it by exactly one factor of frequency,
\begin{equation}
	\chi''(\mathbf{K},\omega) \;=\; \pi\beta\hbar\,\omega\,S_{\rm d}(\mathbf{K},\omega) ,
	\label{eq:fdt}
\end{equation}
which is the classical fluctuation--dissipation theorem. The two are therefore not the same curve: $\chi''$ vanishes linearly at $\omega=0$ and peaks at $\omega=\alpha$, whereas $S_{\rm d}$ peaks at $\omega=0$.


Spin-echo records $I(\mathbf{K},t)$ and returns $F$ and $\alpha$ from the amplitude and decay of the fitted exponential; an energy-resolved measurement obtains the same pair as the area and half width of the diffusive peak; the response to a weak, slow modulation of the coverage probes $\chi(\mathbf{K},\omega)$. The single-exponential description holds over the frequency range these measurements resolve, but not in the far wings, where a Lorentzian has a divergent second moment and the line shape is set by the sub-jump motion left outside this coarse-grained treatment. 

Finally, the two measured functions $F(\mathbf{K})$ and $\alpha(\mathbf{K})$ are not independent. Their product is free of structure, keeping none of the peak-and-dip $\mathbf{K}$ dependence that the interaction imprints on $F$ and on $\alpha$ separately, and varying only through the single-jump geometry,
\begin{equation}
	F(\mathbf{K})\,\alpha(\mathbf{K}) \;=\; \Gamma_{\rm eff}(\theta,T)\,\big[1-f_{\mathbf{K}}\big],
	\label{eq:product}
\end{equation}
the single-adsorbate (Chudley--Elliott) rate, dressed by the interaction only
through the scalar $\Gamma_{\rm eff}$. This is the sum rule
\eqref{eq:sumrule} expressed in the two functions a measurement reports, and
it is exact. 
Any enhancement of the diffusive intensity by short-range correlations is accompanied by an exactly proportional narrowing of the line, and any suppression by a proportional broadening. At small $\mathbf{K}$ the line is motionally narrowed to $\alpha\to\tfrac12 D_cK^{2}$; it is narrowest near a maximum of $F(\mathbf{K})$ and broadens in the flanks, where $F$ falls below its ideal-layer value. Since $F$, $\alpha$ and their product are fixed by the equilibrium layer, the
whole line shape is controlled by a single parameter, which is what
allows the theory to be tested across an entire data set, not at a single
point in the next Section.

\section{Results}
\label{sec:results}

\subsection{Lattice-gas model and short-range correlations}
\label{sec:model}

We treat the adsorbate layer as a lattice gas, that is, as the Ising model written in the variables natural to diffusion. Under the correspondence $s_{\mathbf{R}}=2n_{\mathbf{R}}-1$ \cite{Baxter1982} the spin becomes the site occupation $n_{\mathbf{R}}\in\{0,1\}$, the Ising exchange coupling  becomes the pairwise adsorbate interaction, and the magnetic field the chemical potential $\mu$, conjugate to the number of adsorbates. Two adsorbates on neighbouring sites then form a bond of energy $-\varepsilon$, so $\varepsilon>0$ is attraction and $\varepsilon<0$ repulsion, and summing over bonds gives the grand-canonical Hamiltonian
\begin{equation}
	H \;=\; -\varepsilon \sum_{\langle \mathbf{R}\mathbf{R}'\rangle} n_{\mathbf{R}} n_{\mathbf{R}'} \;-\; \mu \sum_{\mathbf{R}} n_{\mathbf{R}} ,
	\label{eq:latticegas}
\end{equation}
the first sum running over distinct bonds, with $\beta=1/k_BT$ and $z$ the coordination number: $z=2$ for the 1D-channel, $z=4$ for the square lattice, $z=6$ for the triangular lattice of fcc(111) substrates, and so on. The chemical potential is set by $\avg{n_{\mathbf{R}}}=\theta\in(0,1)$, and everything below is parameterized by the pair $(\theta,\beta\varepsilon)$.

Our discussion is based on the pair distribution $P_{\mathbf{L}}$ that builds the SSF \eqref{eq:Fdef}, and the NN occupation statistics that \emph{dress} the hop rate. Along a 1D channel, the relevant geometry when $\mathbf{K}$ is aligned with a row of adsorption sites \cite{TorresMiyares2026c}, the transfer matrix gives both exactly. On a ring of $N$ sites, sharing each single-site term of the grand-canonical weight between the two bonds meeting at that site gives one identical factor per bond,
\begin{equation}
	\exp\!\Big[\beta\varepsilon\sum_i n_i n_{i+1}+\beta\mu\sum_i n_i\Big]
	\;=\; \prod_i T_{n_i,n_{i+1}},
	\qquad
	T_{n,n'} \;=\; e^{\,\beta\varepsilon\,n n'+\frac{\beta\mu}{2}(n+n')},
	\label{eq:Tdef}
\end{equation}
which defines the transfer matrix $T$. In the two-state occupation basis $n=0,1$ (empty and occupied, respectively) the matrix and its eigenvalues are
\begin{equation}
	T=\begin{pmatrix}1 & e^{\beta\mu/2}\\[2pt] e^{\beta\mu/2} & e^{\beta(\varepsilon+\mu)}\end{pmatrix},
	\qquad
	\lambda_\pm=\tfrac12\Big[\,1+e^{\beta(\varepsilon+\mu)}\pm\sqrt{\big(1-e^{\beta(\varepsilon+\mu)}\big)^{2}+4\,e^{\beta\mu}}\,\Big].
	\label{eq:eigenvalues}
\end{equation}
The grand partition function is $\Xi=\mathrm{Tr}\,T^{N}$, and any thermal average is calculated by inserting site operators into that product. With $\hat n=\mathrm{diag}(0,1)$, two sites separated by $r$ spacings give a factor $T^{r}$ between them and $T^{N-r}$ around,
\begin{equation}
	\avg{n_0\,n_r} \;=\; \frac{\mathrm{Tr}\big(\hat n\,T^{r}\,\hat n\,T^{N-r}\big)}{\mathrm{Tr}\,T^{N}} .
	\label{eq:corr-trace}
\end{equation}
$T$ is real and symmetric, so $T^{m}=\sum_{k}\lambda_k^{m}|k\rangle\langle k|$
in its orthonormal eigenbasis. The radical in \eqref{eq:eigenvalues} is strictly
positive, so the two eigenvalues are distinct and $\lambda_+>|\lambda_-|$, and
the eigenvector $|{+}\rangle$ belonging to $\lambda_+$ has components of the
same sign. 
Hence $(\lambda_-/\lambda_+)^{N-r}$ vanishes as $N\to\infty$.
Inserting the spectral decomposition twice turns \eqref{eq:corr-trace} into a double sum over eigenstates, and in the thermodynamic limit $(\lambda_j/\lambda_+)^{N-r}$ kills $j=-$, while $(\lambda_k/\lambda_+)^{r}$ keeps both contributions $k$:
\begin{equation}
	\avg{n_0\,n_r} \;=\; \sum_{k=\pm}\Big(\frac{\lambda_k}{\lambda_+}\Big)^{r}\big|\langle{+}|\hat n|k\rangle\big|^{2}
	\;=\; \theta^{2}+\Big(\frac{\lambda_-}{\lambda_+}\Big)^{r}\big|\langle{+}|\hat n|{-}\rangle\big|^{2},
	\label{eq:two-term}
\end{equation}
the $k=+$ term being $\theta^{2}$ because
$\langle{+}|\hat n|{+}\rangle=\theta$. Measuring each occupation from its mean
as in \eqref{eq:deltan} and subtracting $\theta^{2}$ from \eqref{eq:two-term} leaves the connected correlation, a single power of the eigenvalue ratio, whose amplitude is fixed at $r=0$ by the equal-site variance $\avg{\delta n_0^{2}}=\theta-\theta^{2}$ (using $n_0^{2}=n_0$); hence $|\langle{+}|\hat n|{-}\rangle|^{2}=\theta(1-\theta)$ and the correlation decays geometrically,
\begin{equation}
	\avg{\delta n_0\,\delta n_r} \;=\; \theta(1-\theta)\,\eta^{|r|},
	\qquad
	\eta \;\equiv\; \frac{\lambda_-}{\lambda_+}\in(-1,1),
	\label{eq:geometric}
\end{equation}
governed by the single ratio $\eta$ of the two transfer-matrix eigenvalues. Since $\lambda_+\lambda_-=\det T=e^{\beta\mu}(e^{\beta\varepsilon}-1)$, the smaller eigenvalue is negative whenever the interaction is repulsive, so $\eta<0$ makes \eqref{eq:geometric} alternate in sign from site to site. 

However, $\mu$ is not a laboratory variable. One should then write equations in terms of the coverage, which is tied to $\mu$ by the equation of state, with $\Xi\to\lambda_+^{N}$ as $N\to\infty$:
\begin{equation}
	\theta \;=\; \frac{1}{N}\,\frac{\partial \ln\Xi}{\partial(\beta\mu)}
	\;\xrightarrow[N\to\infty]{}\; \frac{\partial \ln\lambda_+}{\partial(\beta\mu)} ,
	\label{eq:eos}
\end{equation}
a monotonically increasing relation $\theta(\mu)$ at fixed
$\beta\varepsilon$. Instead of inverting it, one can reach
$\eta(\theta,\beta\varepsilon)$ by a route on which $\mu$ never appears. Writing $\psi$ for the normalized eigenvector $|{+}\rangle$, the joint probability of occupations $n$ and $n'$ on two adjacent sites is $p_{nn'}=\psi_n T_{nn'}\psi_{n'}/\lambda_+$ in the thermodynamic limit; summing over $n'$ returns the single-site marginal $\psi_n^{2}$, so that $\psi_1^{2}=\theta$. In the cross ratio the eigenvector components and $\lambda_+$ cancel in pairs, leaving a quantity fixed by the matrix elements alone, in which $\mu$ cancels as well:
\begin{equation}
	\frac{p_{11}\,p_{00}}{p_{01}\,p_{10}} \;=\; \frac{T_{11}T_{00}}{T_{01}T_{10}} \;=\; e^{\beta\varepsilon} .
	\label{eq:quasichem}
\end{equation}
This is exact for the chain, and it is a law of mass action for the exchange
$11+00\rightleftharpoons01+10$, with $e^{\beta\varepsilon}$ as equilibrium
constant. Its natural variable is not the correlation but the pair probability
itself. Writing $x\equiv p_{01}=p_{10}$ for the chance that two adjacent sites
are one occupied and one empty, the marginals $p_{11}+p_{01}=\theta$ and
$p_{00}+p_{01}=1-\theta$ fix the other two, $p_{11}=\theta-x$ and
$p_{00}=1-\theta-x$, so that \eqref{eq:quasichem} becomes
$(\theta-x)(1-\theta-x)=x^{2}e^{\beta\varepsilon}$, a quadratic in $x$,
\begin{equation}
	\big(1-e^{\beta\varepsilon}\big)\,x^{2} \;-\; x \;+\; S_0 \;=\; 0 ,
	\qquad S_0\equiv\theta(1-\theta) ,
	\label{eq:etaquad}
\end{equation}
whose discriminant is $\zeta^{2}=1-4S_0(1-e^{\beta\varepsilon})$. The root that
returns $x\to S_0$ at $\varepsilon=0$ is
$x=(1-\zeta)/2(1-e^{\beta\varepsilon})$, and since the connected correlation
\eqref{eq:geometric} makes $p_{01}=S_0(1-\eta)$, the closed form is $\eta=1-x/S_0$,
\begin{equation}
	\eta(\theta,\beta\varepsilon) \;=\; \frac{\zeta-1}{\zeta+1},
	\qquad
	\zeta \;\equiv\; \sqrt{\,1-4\theta(1-\theta)\big(1-e^{\beta\varepsilon}\big)\,},
	\label{eq:eta}
\end{equation}
a single dimensionless number carrying the entire correlation: $\eta=0$ without interaction, $\eta>0$ for attraction, $\eta<0$ for repulsion. Since $\avg{n_0 n_1}=\theta^{2}+\theta(1-\theta)\eta$, it measures the excess or depletion of NN occupation relative to $\theta$. 
Two limits follow at once: $\eta=\tanh(\beta\varepsilon/4)$ at half coverage, and $\eta\simeq\theta(1-\theta)\,\beta\varepsilon$ at weak coupling. Its magnitude grows as the correlations strengthen, and reaches $|\eta|\to1$ only under very strong coupling. This is the limit that bounds the theory developed here, which throughout treats a weakly-to-moderately correlated, spatially uniform adlayer.
Where that limit sits can be stated precisely. The radicand of $\zeta$ is positive for every $(\theta,\beta\varepsilon)$: for attraction $1-e^{\beta\varepsilon}<0$ and $\zeta>1$, while for repulsion $4\theta(1-\theta)\le1$ and $1-e^{\beta\varepsilon}<1$, so the product never reaches unity. Hence $\zeta$ is real and $|\eta|<1$ strictly at every coverage and coupling: the chain has no transition and its correlation length stays finite. The strong-coupling caveat is therefore not about a singularity of \eqref{eq:eta} but about the accuracy of the two-dimensional closures that inherit it.

All of this is exact in one dimension. On the two-dimensional lattice we keep the same local $\eta(\theta,\beta\varepsilon)$ as the NN correlation (the pair approximation \cite{Pelizzola2005}) and obtain the momentum dependence of the structure factor from a collective (random-phase) treatment. The extension is an approximation: the true NN correlation of the square or triangular lattice is not \eqref{eq:eta}, which knows nothing of the closed loops absent from a chain.

\subsection{The static structure factor}
\label{sec:ssf}

We first analyze it exactly along a 1D channel, where the transfer matrix gives a lattice Lorentzian, and then at collective (random-phase) level for the full surface and for interactions of any range.

Along the 1D-channel, a separation of $n$ cells carries a phase $Kan\cos\tilde\beta$, and conditioning \eqref{eq:geometric} on the origin being occupied gives
\begin{equation}
	P_n \;=\; \theta+(1-\theta)\,\eta^{|n|},
	\label{eq:Pn}
\end{equation}
with $P_0=1$ and $P_n\to\theta$ far away, as it should be. Its uniform background $\theta$ and decaying correlation $(1-\theta)\eta^{|n|}$ transform into the two parts of the $t=0$ signal, $F(\mathbf{K})=A(\mathbf{K})+B(\mathbf{K})$:
\begin{equation}
	F(\mathbf{K}) \;=\; \underbrace{\theta \sum_{n} e^{\,iKan\cos\tilde\beta}}_{A(\mathbf{K})}
	\;+\;\underbrace{(1-\theta)\sum_{n}\eta^{|n|}e^{\,iKan\cos\tilde\beta}}_{B(\mathbf{K})}.
	\label{eq:FK-split}
\end{equation}
As mentioned above, the first term is the Dirac comb at the Bragg points, the elastic part, carrying no dynamics; the second holds all the interaction dependence. Writing $x\equiv\Keff$ and separating the $n=0$ term from the two tails, one has
\begin{equation}
	\sum_{n=-\infty}^{\infty}\eta^{|n|}e^{ixn}
	\;=\; 1+\sum_{n\ge1}(\eta e^{ix})^{n}+\sum_{n\ge1}(\eta e^{-ix})^{n}
	\;=\; \frac{1-\eta^{2}}{1-2\eta\cos x+\eta^{2}},
	\label{eq:geo-sum}
\end{equation}
each tail convergent because $|\eta|<1$. Away from the Bragg points the comb $A(\mathbf{K})$ vanishes and only the diffusive part survives, so $F(\mathbf{K})=B(\mathbf{K})$ at every momentum an experiment actually samples. In terms of the  parameters the experiment controls,
\begin{equation}
	F(\mathbf{K}) \;=\; (1-\theta)\,\frac{1-\eta^{2}}{1-2\eta\cos(\Keff)+\eta^{2}} .
	\label{eq:FK}
\end{equation}
Its shape comes from the geometric decay $\avg{\delta n_0\,\delta n_r}\propto|\eta|^{|r|}$. Written as an exponential in the separation $ra$, that decay defines the correlation length, $\xi$,
\begin{equation}
	|\eta|^{|r|}\;=\;e^{-\,|r|\,a/\xi}
	\quad\Longrightarrow\quad
	\frac{\xi}{a}\;=\;-\frac{1}{\ln|\eta|},
	\label{eq:corrlength}
\end{equation}
growing without bound as $|\eta|\to1$. For attraction, the diffusive intensity peaks at the zone centre; expanding \eqref{eq:FK} for small $x$ turns the denominator into $(1-\eta)^{2}+\eta\,x^{2}$, so
\begin{equation}
	F(\mathbf{K}) \;\simeq\; \frac{(1-\theta)(1-\eta^{2})}{(1-\eta)^{2}+\eta\,x^{2}}
	\;\simeq\; \frac{F_{\rm pk}}{1+\big(x\,\xi_{\rm OZ}/a\big)^{2}},
	\qquad
	F_{\rm pk}\equiv F(\mathbf{0}) ,
	\qquad
	\frac{\xi_{\rm OZ}}{a}\;=\;\frac{\sqrt{|\eta|}}{1-|\eta|} ,
	\label{eq:lorentz}
\end{equation}
with $x$ now measured from the peak. The Ornstein--Zernike length
$\xi_{\rm OZ}$ governing the line shape is not the exponential-decay length \eqref{eq:corrlength} governing the correlation itself; the two coincide only as $|\eta|\to1$, where $\ln|\eta|\simeq-(1-|\eta|)$. The intensity halves at $x=\pm(1-|\eta|)/\sqrt{|\eta|}$, so the full width at
half maximum is twice that, tending to $2/\xi$ as the correlations
lengthen. Written in $|\eta|$, \eqref{eq:lorentz} and that width cover both
signs of the interaction at once: for repulsion, the same expansion about the zone boundary (ZB), with $x=\pi+y$, turns the denominator into $(1-|\eta|)^{2}+|\eta|y^{2}$ and gives the mirror-image peak there, of the same width measured from the ZB. Equation \eqref{eq:FK} is what the literature calls a lattice Lorentzian, the periodic form of the Ornstein--Zernike line shape of a correlated fluid \cite{Hansen2013}. Its two high-symmetry values,
\begin{equation}
	F(\mathbf{0})=(1-\theta)\,\frac{1+\eta}{1-\eta},
	\qquad
	F_{\rm ZB}=(1-\theta)\,\frac{1-\eta}{1+\eta},
	\label{eq:FKzone}
\end{equation}
are reciprocals of one another about the ideal value $1-\theta$: attraction piles intensity at the zone centre and depletes the boundary, repulsion does the reverse. The zone-centre value is proportional to the isothermal compressibility \cite{Hansen2013}.

The transfer matrix works only in one dimension, where the partition function factorizes bond by bond; neither the surface nor a longer-ranged interaction allows that. For these, we compute $F(\mathbf{K})$ from the density fluctuations at Gaussian level, the random-phase approximation \cite{ChaikinLubensky,Hansen2013}. The analysis proceeds through the density wave of wavevector $\mathbf{K}$, of
complex amplitude \eqref{eq:densitywave}.
As mentioned above, the mean-square value per site is the density--density structure factor, $S(\mathbf{K})=\avg{|\delta n_{\mathbf{K}}|^{2}}/N$ \cite{ChaikinLubensky}. Expanding the square shows that it is the lattice Fourier transform of the density-fluctuation correlation,
\begin{equation}
	S(\mathbf{K})=\sum_{\mathbf{L}}\avg{\delta n_0\,\delta n_{\mathbf{L}}}\,e^{\,i\mathbf{K}\cdot\mathbf{L}}
	=\sum_{\mathbf{L}}\big(\avg{n_0 n_{\mathbf{L}}}-\theta^{2}\big)\,e^{\,i\mathbf{K}\cdot\mathbf{L}} .
\end{equation}
The subtracted constant contributes only a Bragg peak, and away from the Bragg
points $S$ and $F$ differ only by the factor $\theta$ of \eqref{eq:FSrel}.
For the ideal adlayer $S(\mathbf{K})$ reduces to the site variance $S_0\equiv\theta(1-\theta)$ and $F(\mathbf{K})$ to the Langmuir value $1-\theta$.

The mean-field value of $S(\mathbf{K})$ rests on one idea: each density wave is
an independent thermally excited mode whose free energy grows quadratically
with its amplitude,
\begin{equation}
	\Delta\mathcal{F} \;=\; \sigma(\mathbf{K})\,\frac{|\delta n_{\mathbf{K}}|^{2}}{2N} ,
	\label{eq:freeenergy}
\end{equation}
the factor $N$ making $\Delta\mathcal{F}$ extensive in the same way as $S$.
Equipartition then fixes the mean-square amplitude of a classical quadratic
mode, $S(\mathbf{K})=\avg{|\delta n_{\mathbf{K}}|^{2}}/N=k_BT/\sigma(\mathbf{K})$,
so that everything reduces to identifying $\sigma(\mathbf{K})$, the inverse of
the density response of the layer at wavevector $\mathbf{K}$, and at
$\mathbf{K}=0$ the inverse compressibility met in \eqref{eq:Dhydro}.

Without interaction, the sites are statistically
independent and $S(\mathbf{K})=S_0$ at every momentum, which by the same equipartition makes $\sigma$ purely entropic, $k_BT/S_0$,
and the same at every wavelength. The interaction adds the free energy of the
modulation itself, which for a pair energy $v(\mathbf{R})$ between sites at
separation $\mathbf{R}$ is its lattice transform. Hence
\begin{equation}
	\sigma(\mathbf{K}) \;=\; \frac{k_BT}{S_0}\;+\;v(\mathbf{K}),
	\qquad
	v(\mathbf{K})=\sum_{\mathbf{R}}v(\mathbf{R})\,e^{\,i\mathbf{K}\cdot\mathbf{R}} ,
	\label{eq:stiffness}
\end{equation}
so that $S(\mathbf{K})=S_0/[1+\beta S_0 v(\mathbf{K})]$, the random-phase
(Ornstein--Zernike) structure factor~\cite{ChaikinLubensky,Hansen2013}.
Dividing by $\theta$ as in \eqref{eq:FSrel} gives the diffusive amplitude in
closed form,
\begin{equation}
	F(\mathbf{K}) \;=\; \frac{1-\theta}{1+\beta\,\theta(1-\theta)\,v(\mathbf{K})}.
	\label{eq:FKrpa}
\end{equation}
The sign of the interaction is transparent in the denominator: where a density wave raises the free energy, $v(\mathbf{K})>0$, the mode is
stiffer and $F<1-\theta$; where it is favoured, the mode is softer and $F>1-\theta$. For the NN lattice
gas \eqref{eq:latticegas}, a site couples with energy $-\varepsilon$ to each of its $z$ neighbours, so
\begin{equation}
	v(\mathbf{K}) \;=\; -\varepsilon\sum_{\boldsymbol\delta}e^{\,i\mathbf{K}\cdot\boldsymbol\delta}
	\;=\; -\varepsilon\,z\,\gamma(\mathbf{K}),
	\qquad
	\gamma(\mathbf{K})=\frac{1}{z}\sum_{\boldsymbol\delta}e^{\,i\mathbf{K}\cdot\boldsymbol\delta},
	\label{eq:vtilde-nn}
\end{equation}
where $\boldsymbol\delta$ runs over the $z$ neighbour vectors and $\gamma(\mathbf{K})$ is their average phase, equal to $1$ at the zone centre and $-1$ at a bipartite zone boundary. The
qualifier matters: the triangular lattice is not bipartite and there $\gamma$
bottoms out at $-\tfrac12$, so a repulsive layer on fcc(111) is frustrated and
its zone-boundary enhancement weaker than the square-lattice case would
suggest.
 Substituting into \eqref{eq:FKrpa},
\begin{equation}
	F(\mathbf{K}) \;=\; \frac{1-\theta}{1-\beta\varepsilon\,z\,\theta(1-\theta)\,\gamma(\mathbf{K})}.
	\label{eq:FK2d}
\end{equation}
This has the same functional form as the exact one-dimensional result \eqref{eq:FK}, peaked at the zone centre for attraction, at the zone boundary for repulsion and flat when $\varepsilon=0$; both belong to the family $1/(A-B\cos x)$, but with $A=(1+\eta^{2})/(1-\eta^{2})$ in \eqref{eq:FK} against $A=1$ here, so at $z=2$ the two agree only to leading order in $\beta\varepsilon$, through $\eta\simeq\theta(1-\theta)\beta\varepsilon$.

Two points deserve emphasis at this level. The denominator reaches zero when $\beta\varepsilon z\theta(1-\theta)\gamma(\mathbf{K}^\star)=1$ at the favoured wavevector; the divergence is not physical but marks where the Gaussian treatment fails. That it is an artefact of the closure, not a feature of the layer, is shown by the one-dimensional case, where the exact \eqref{eq:eta} keeps $|\eta|<1$ at every coverage and coupling and \eqref{eq:FK} never diverges. And the mean-field form overestimates the growth of correlations, so away from weak coupling the true $F$ lies between \eqref{eq:FK2d} and \eqref{eq:FK}: we use the exact Lorentzian along a diffusion channel and the collective form on the surface. The collective form needs only $v(\mathbf{K})$, so it holds for an interaction of any range --- the right tool for adsorbates on metal surfaces, where conduction electrons mediate a long-ranged interaction whose sign oscillates with separation \cite{Repp2000}. Such a tail enters only through $v(\mathbf{K})$ in \eqref{eq:FKrpa}, leaving every step above unchanged.

\subsection{The hop rate and the closed form}
\label{sec:dyn}

Dynamics enters through the hop rate. An adsorbate exchanges with a NN vacant site in Kawasaki dynamics \cite{Kawasaki1966}, which conserves adsorbate number as it should. The interaction changes $\Gamma_0$ in two ways, both fixed by the NN statistics carried by $\eta$. From $\avg{n_0 n_1}=\theta^{2}+\theta(1-\theta)\eta$, the neighbour of an occupied site is occupied or empty with probability
\begin{equation}
	q\equiv P(1|1)=\theta+(1-\theta)\,\eta,
	\qquad
	P(0|1)=1-q=(1-\theta)(1-\eta).
	\label{eq:conditional}
\end{equation}
First, the hop requires an empty target, available with probability not the bare $1-\theta$ but the correlated $P(0|1)=(1-\theta)(1-\eta)$, which attraction lowers and repulsion raises. Second, the adsorbate must break the bonds to its other occupied neighbours. If $\nu$ of the remaining $z-1$ neighbours are occupied, the barrier rises by $\nu\varepsilon$; averaging $e^{-\beta\varepsilon\nu}$ over the binomial distribution of $\nu$, each neighbour occupied with probability $q$, gives
\begin{equation}
	\avg{e^{-\beta\varepsilon\nu}} \;=\; \big[\,1-q+q\,e^{-\beta\varepsilon}\big]^{\,z-1}.
	\label{eq:mgf}
\end{equation}
The effective hop rate is the bare rate times these two factors,
\begin{equation}
	\Gamma_{\rm eff}(\theta,T) \;=\; \Gamma_0\,(1-\theta)(1-\eta)\,\big[\,1-q+q\,e^{-\beta\varepsilon}\big]^{\,z-1},
	\label{eq:gamma-eff}
\end{equation}
valid for any lattice, $z$ being the coordination number at hand. Only the bonds broken on leaving enter, the transition state being taken as non-interacting.
At $\varepsilon=0$, it returns the Langmuir site-blocking rate $\Gamma_0(1-\theta)$; attraction suppresses it, both trapping the adsorbate and blocking its target, and repulsion enhances it. The interaction renormalizes the dynamics, not the geometry: in a uniform adlayer every neighbour direction stays equivalent, so $w_{\mathbf{L}}$ and $f_{\mathbf{K}}$ are unchanged.
Equation \eqref{eq:gamma-eff} takes its two factors as separate averages, as though the emptiness of the target site and the occupation of the remaining neighbours were independent of one another. They are not, in general.
The quantity that governs the dynamics is the mean escape rate of an adsorbate, and that is a single equilibrium average of the configuration-dependent rate together with the constraints the move carries --- the departure site $i$ occupied, the target $j$ empty --- so that $\Gamma_{\rm eff}=(z/\theta)\avg{w_{i\to j}\,n_i(1-n_j)}$, the average running over the Gibbs measure of \eqref{eq:latticegas} and $\theta$ converting a rate per site into a rate per adsorbate. Along a channel, the rate $w_{i\to j}$ depends on the single remaining neighbour of $i$, so this is an average over three consecutive sites: the departure site, the target, and the neighbour whose bond must be broken. The transfer matrix evaluates it in closed form, and the result reproduces \eqref{eq:gamma-eff} at every coverage and coupling.
On the surface, the closed loops destroy that conditional independence and \eqref{eq:gamma-eff} becomes a pair approximation, accurate at weak coupling and in the dilute and near-saturation wings. Replacing the fluctuating rate by its equilibrium mean is the standard mean-rate treatment of lattice-gas transport \cite{Reed1981,Gomer1990}; by \eqref{eq:sumrule} that replacement is exact for the local rate of \eqref{eq:eom}, the mean being the only feature of the fluctuating rate that the first derivative of the signal can see. What the mean-rate treatment does discard is the memory function, and that is retained here.
The rates obey detailed balance with respect to \eqref{eq:latticegas} and the statics $\eta,q$ dressing $\Gamma_{\rm eff}$ come from the Gibbs measure that fixes $F(\mathbf{K})$, so statics and dynamics share one thermodynamic footing, which lets the Darken relation emerge below instead of being imposed.

With a Markovian process of hop rate $\Gamma_{\rm eff}$ and an unchanged single-jump CF, the compound-Poisson result \eqref{eq:ce} applies unchanged to one adsorbate diffusing in the interacting adlayer,
\begin{equation}
	I(\mathbf{K},t;\theta,T) \;=\; \exp\!\big[-\alpha_1(\mathbf{K};\theta,T)\,t\big],
	\qquad
	\alpha_1(\mathbf{K};\theta,T) \;=\; \Gamma_{\rm eff}(\theta,T)\,\big[1-f_{\mathbf{K}}\big],
	\label{eq:isf-one}
\end{equation}
a pure exponential whose rate $\alpha_1$ carries the entire interaction effect
through the scalar $\Gamma_{\rm eff}$; for NN hops along a channel, the Chudley--Elliott shape with a \emph{dressed} prefactor.
Equation \eqref{eq:isf-one} follows one adsorbate at the mean-rate, or \emph{jump},
level. The coherent experiment instead measures the relaxation of the collective
density, whose amplitude and local rate were fixed in Sec.~\ref{sec:coeff} without
reference to any model: the amplitude by the identity $I(\mathbf{K},0)=F(\mathbf{K})$
and the rate by the sum rule \eqref{eq:sumrule}, in which the exchange current
\eqref{eq:current} enters through $\Gamma_{\rm eff}=zJ/\theta$. The closed form \eqref{eq:gamma-eff} for $\Gamma_{\rm eff}$ and
\eqref{eq:FK}--\eqref{eq:FK2d} for the structure factor turn \eqref{eq:sumrule} into an elementary function of momentum
transfer, coverage and temperature, and discarding the memory function then gives
the whole ISF,
\begin{equation}
	I(\mathbf{K},t) \;=\; F(\mathbf{K})\;
	\exp\!\left[-\,\frac{\Gamma_{\rm eff}\,\big[1-f_{\mathbf{K}}\big]}{F(\mathbf{K})}\;t\right],
	\label{eq:main}
\end{equation}
every ingredient being closed-form: $F$ from Eqs.~\eqref{eq:eta}--\eqref{eq:FK2d}, $\Gamma_{\rm eff}$ from \eqref{eq:gamma-eff}, and $f_{\mathbf{K}}$ from the jump geometry. The dephasing rate
\begin{equation}
		\alpha(\mathbf{K}) \;=\; \frac{\Gamma_{\rm eff}\,\big[1-f_{\mathbf{K}}\big]}{F(\mathbf{K})}
	\label{eq:alpha}
\end{equation}
is the local rate \eqref{eq:sumrule} of Sec.~\ref{sec:coeff}, now an elementary
function of $(\theta,\beta\varepsilon)$; it is the one-adsorbate rate
\eqref{eq:isf-one} divided by the SSF, so relaxation slows wherever the adlayer has built up structure ($F$ large) and quickens where structure is suppressed. This narrowing of the corresponding line shape by the inverse structure factor is the coherent, CF counterpart of the de Gennes narrowing of dense-fluid scattering, where the coherent linewidth scales as $1/S(\mathbf{K})$ \cite{deGennes1959}; the factorized form \eqref{eq:main} is the lattice-gas, closed-form realization of that narrowing.
Compared to the single-adsorbate result \eqref{eq:ce}, \eqref{eq:main} is that same exponential with both of its coefficients renormalized: the amplitude from unity to $F(\mathbf{K})$, and the rate from $\Gamma_0[1-f_{\mathbf{K}}]$ to $\Gamma_{\rm eff}[1-f_{\mathbf{K}}]/F(\mathbf{K})$, each renormalization carried by $\eta$ alone. Since the fit \eqref{eq:diffusive} performed on the data has exactly this form, the theory speaks directly to the two numbers an experiment reports, and it is those two numbers, not the shape, that carry its content.


Equation \eqref{eq:main} contains no adjustable parameter: $F$, $\Gamma_{\rm eff}$ and $f_{\mathbf{K}}$ are each fixed by $(\theta,\beta\varepsilon)$ through $\eta$ and by the jump geometry, so the construction is falsifiable, not tunable. It is also exact on two entire boundaries of the parameter plane: at $\varepsilon=0$, for every coverage and momentum, 
and again as $\theta\to0$, where every dressing factor tends to unity. These are precisely the boundaries on which $\eta$ vanishes, and $\eta$ is the sole carrier of the interaction, so the closed form departs from exactness continuously and only as the correlations build. At $\varepsilon=0$, the statement is stronger than a limit: by \eqref{eq:sep}, the density wave is then an exact relaxation mode at every momentum transfer, $\Delta(\mathbf{K})$ vanishes identically, and \eqref{eq:main} is not merely accurate but exact at every time. The sharpest test is not internal, however, but the one performed in Sec.~\ref{sec:nacu}: \eqref{eq:main} forces the product $F(\mathbf{K})\alpha(\mathbf{K})$ to be free of structure, and published spin-echo data satisfy that constraint point by point on a layer whose structure factor is not the Ising one.

\begin{figure}[!tbp]
	\centering
	\includegraphics[width=0.99\textwidth]{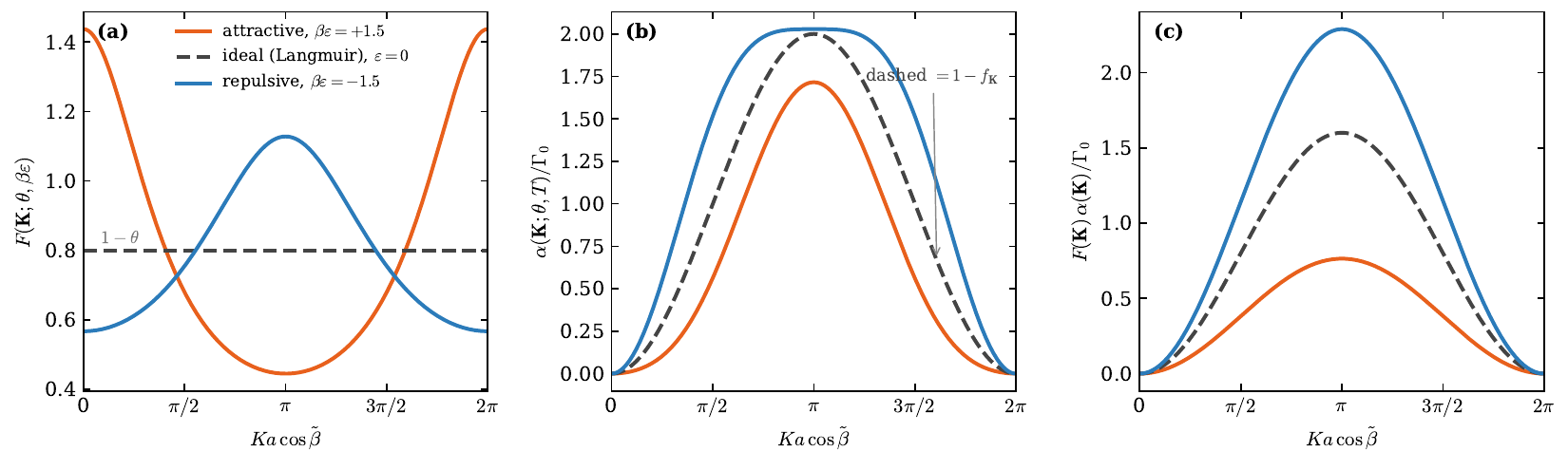}
	\caption{Analytic ingredients of the interacting ISF, Eq.~\eqref{eq:main}, for a one-dimensional channel ($z=2$) at coverage $\theta=0.2$: attractive ($\beta\varepsilon=+1.5$, $\eta=+0.28$), ideal ($\varepsilon=0$, dashed) and repulsive ($\beta\varepsilon=-1.5$, $\eta=-0.17$) adlayers. (a) Static structure factor $F(\mathbf{K})$, Eq.~\eqref{eq:FK}: attraction gathers diffuse intensity at the zone centre, repulsion at the zone boundary, while the ideal adlayer is flat at $1-\theta$. (b) Dephasing rate $\alpha(\mathbf{K})/\Gamma_0$, Eq.~\eqref{eq:alpha}. The dashed curve is at once the ideal adlayer and the bare single-jump shape $1-f_{\mathbf{K}}=1-\cos(\Keff)$: at $\varepsilon=0$ the site-blocking factor cancels between $\Gamma_{\rm eff}$ and $F$, so the two coincide identically. Away from that point the interaction shifts the rate through $\Gamma_{\rm eff}$ --- to $0.38\,\Gamma_0$ for attraction and $1.14\,\Gamma_0$ for repulsion, against $0.80\,\Gamma_0$ for the ideal layer --- and distorts its shape through $1/F(\mathbf{K})$, flattening the repulsive curve around the structure peak at the zone boundary. (c) The two effects separated: the product $F(\mathbf{K})\alpha(\mathbf{K})/\Gamma_0$, Eq.~\eqref{eq:product}, retains no trace of the peak and dip of panels (a) and (b) and reduces to $1-f_{\mathbf{K}}$ scaled by $\Gamma_{\rm eff}$ alone. This structureless product is the prediction tested against data in Sec.~\ref{sec:nacu}.}
	\label{fig:main}
\end{figure}

The shape of $F(\mathbf{K})$ in Fig.~\ref{fig:main} is set by the sign of $\eta$, flat at the Langmuir value $1-\theta$ for the ideal adlayer, the finite-coverage generalization of the structureless ISA amplitude \cite{MartinezCasado2007,TorresMiyares2026c}. Panels (a) and (b) are mirror images, and panel (c) shows what survives their product: $F(\mathbf{K})\,\alpha(\mathbf{K})=\Gamma_{\rm eff}[1-f_{\mathbf{K}}]$ carries no structure at all, the interaction entering only as the scalar $\Gamma_{\rm eff}$. Four exact limits connect the closed form to results known independently. At $t=0$, the identity $I(\mathbf{K},0)=F(\mathbf{K})$ makes the measured amplitude the SSF. At $\varepsilon=0$, and any coverage, $\eta=0$, $F=1-\theta$ and $\Gamma_{\rm eff}=\Gamma_0(1-\theta)$, so \eqref{eq:main} collapses to $I(\mathbf{K},t)=(1-\theta)\exp[-\Gamma_0(1-f_{\mathbf{K}})t]$: the site-blocking factor appears once in $\Gamma_{\rm eff}$ and once in $F$, and the two cancel, so the coherent relaxation proceeds at the bare rate even though each hop is blocked. This is the symmetric exclusion process, whose chemical diffusion coefficient stays at $D_0$ for all coverages \cite{Kutner1981}, recovered without being tuned to it. As $\mathbf{K}\to0$ the signal reduces to Fickian relaxation, $\alpha\to\tfrac12 D_c K^{2}$ with $D_c$ fixed by $F(\mathbf{0})$.
This limit is exact as a statement about the initial rate. And as $\theta\to0$ every dressing factor tends to unity and \eqref{eq:main} returns the single-adsorbate theory of Ref.~\cite{TorresMiyares2026c} exactly.

\subsection{Transport coefficients}
\label{sec:transport}

No separate transport calculation is needed: the diffusion coefficients are already inside \eqref{eq:isf-one} and \eqref{eq:main}, and the defining property of a CF extracts them. Its derivatives at the origin return the moments of $\rho_{\mathbf{L}}(t)$, whose total weight is $I(\mathbf{0},t)$, and the logarithm of the normalized CF generates the cumulants of the
projected displacement given by \eqref{eq:cumulants}. Cumulants   separate the two
contents of the signal additively, the equilibrium structure from the
transport, so that the diffusion coefficient is a pure slope; and for a
Markovian jump sequence it is linear in time at every order, which is what
makes the Fickian limit a result and not an assumption. At second order the
choice is immaterial, the first cumulant vanishing by the symmetry of
Sec.~\ref{sec:cf}.
Forward and backward jumps are equally likely in a uniform adlayer, so the
single-jump CF \eqref{eq:fK} is even in $\mathbf{K}$ and the first cumulant
vanishes: the second cumulant is then the second moment itself,
$\kappa_2(t)=\avg{L_\parallel^{2}(t)}$, the mean-square displacement along
$\mathbf{K}$. That is the first of two facts making the differentiation
immediate; it also makes the expansion of $f_{\mathbf{K}}$ start at second
order,
\begin{equation}
f_{\mathbf{K}} \;=\; 1-\tfrac12 K^{2}\avg{\ell^{2}}+O(K^{4}),
\qquad
\avg{\ell^{2}}\;=\;\sum_j w_j\,a_j^{2}\cos^{2}\tilde\beta_j ,
\label{eq:l2}
\end{equation}
where $\avg{\ell^{2}}$ is the mean-square projected jump length [\eqref{eq:Lpar}]. It depends on the neighbour shell and not on the interaction, and only the
ratios $D_1/D_0$ and $D_c/D_0$ are quoted below, in which it cancels. Second, $1-f_{\mathbf{0}}=0$, so the exponent of \eqref{eq:main} vanishes at the zone centre and $I(\mathbf{0},t)=F(\mathbf{0})$ at all times, as adsorbate-number conservation requires.

Applied to the one-adsorbate ISF \eqref{eq:isf-one}, already normalized, the logarithm $\ln I=-\Gamma_{\rm eff}[1-f_{\mathbf{K}}]t$ is a cumulant generating function in closed form. Inserting \eqref{eq:l2} and differentiating twice gives the mean-square displacement at this mean-rate level,
\begin{equation}
\avg{L_\parallel^{2}(t)} \;=\; \avg{\ell^{2}}\;\Gamma_{\rm eff}(\theta,T)\;t ,
\label{eq:msd-one}
\end{equation}
strictly linear in time at every time, not only asymptotically, because a Markovian jump sequence carries no memory. Its slope is the jump diffusion coefficient\footnote{
In a strictly single-file channel tagged motion is subdiffusive and no constant tracer coefficient exists. The collective $D_c$ is untouched by that statement.} along the observation direction,
\begin{equation}
D_1(\theta,T) \;=\; \avg{\ell^{2}}\;\Gamma_{\rm eff}(\theta,T)
\;=\; D_0\,(1-\theta)(1-\eta)\,\big[1-q+q\,e^{-\beta\varepsilon}\big]^{\,z-1},
\label{eq:Ds}
\end{equation}
with $D_0= \avg{\ell^{2}}\,\Gamma_0$ the isolated-adsorbate value.\footnote{$\Gamma_0$ is the total jump rate, as in Ref.~\cite{TorresMiyares2026c}, and $D=\avg{\ell^{2}}\,\Gamma=\avg{L_\parallel^{2}(t)}/t$ from \eqref{eq:msd-one}, which omits the factor $\tfrac12$ of the Einstein relation, so that $\alpha_1(\mathbf{K}\to0)=\tfrac12 D_1K^{2}$. Only the ratios $D_1/D_0$ and $D_c/D_0$ enter the results, in which this convention cancels.} Higher cumulants come free: $\kappa_{2m}(t)=\Gamma_{\rm eff}\avg{\ell^{2m}}t$ is linear in time for every $m$, so $\kappa_4/\kappa_2^{2}$ decays as $1/t$ and the Fickian reading of \eqref{eq:msd-one} is a consequence, not an assumption.

The same two derivatives applied to \eqref{eq:main} deliver the collective coefficient, the equilibrium structure now contributing too. Since $I(\mathbf{0},t)=F(\mathbf{0})$,
\begin{equation}
\ln I_N(\mathbf{K},t) \;=\; \ln\frac{F(\mathbf{K})}{F(\mathbf{0})}
\;-\;\frac{\Gamma_{\rm eff}\,\big[1-f_{\mathbf{K}}\big]}{F(\mathbf{K})}\;t ,
\label{eq:lnIhat}
\end{equation}
a static term plus a term linear in time, the two never mixing. Differentiating twice at the zone centre, with \eqref{eq:l2} and $F(\mathbf{K})=F(\mathbf{0})+O(K^{2})$ so that the structure factor enters the ratio $[1-f_{\mathbf{K}}]/F(\mathbf{K})$ only at $O(K^{4})$,
\begin{equation}
\kappa_2(t) \;=\; \ell_0^{2} \;+\; D_c\,t ,
\qquad
\ell_0^{2}=-\frac{\partial^{2}\ln F}{\partial K^{2}}\bigg|_{\mathbf{K}=0},
\qquad
D_c \;=\; \frac{\avg{\ell^{2}}\;\Gamma_{\rm eff}}{F(\mathbf{0})} \;=\; \frac{D_1}{F(\mathbf{0})} .
\label{eq:msd-coll}
\end{equation}
The split is the physical content of the coherent measurement. The constant $\ell_0^{2}$ is the mean-square separation already present in the equilibrium layer; for the channel Lorentzian \eqref{eq:FK} it is $\ell_0^{2}=a^{2}\cos^{2}\tilde\beta\;2\eta/(1-\eta)^{2}$, either sign being allowed because this is a cumulant, not of a probability distribution. Being a static offset, it carries no transport; only the second term does, its slope being the chemical diffusion coefficient: $\alpha_1\to\tfrac12 D_1K^{2}$ and $\alpha\to\tfrac12 D_cK^{2}$ as $\mathbf{K}\to0$.

The moment route provides $D_c=D_1/F(\mathbf{0})$ as a result, not a definition, and this is where thermodynamics enters. At infinite wavelength the structure factor measures the variance $\langle(\Delta N)^{2}\rangle$ of the adsorbate number in an open region; mean and fluctuation come from the same partition function, so $S(\mathbf{0})=\partial\theta/\partial(\beta\mu)$ per site --- the compressibility sum rule. With $F=S/\theta$ [\eqref{eq:FSrel}],
\begin{equation}
F(\mathbf{0})=\frac{S(\mathbf{0})}{\theta}=\frac{1}{\theta}\,\frac{\partial\theta}{\partial(\beta\mu)}=\frac{\partial\ln\theta}{\partial(\beta\mu)} .
\label{eq:compress-gen}
\end{equation}
Evaluating the derivative with the transfer-matrix equation of state gives the closed 1D sum rule
\begin{equation}
F(\mathbf{0}) \;=\; \frac{(1-\theta)(1+\eta)}{1-\eta}
\;=\;
\left[\theta\,\frac{\partial(\beta\mu)}{\partial\theta}\right]^{-1}
\;\equiv\; \Theta^{-1}(\theta,T),
\label{eq:compress}
\end{equation}
with $\Theta=\partial(\beta\mu)/\partial\ln\theta$ the thermodynamic factor,
the inverse compressibility of the adlayer. Substituting $F(\mathbf{0})=\Theta^{-1}$ into $D_c=D_1/F(\mathbf{0})$ gives at once
\begin{equation}
D_c(\theta,T) \;=\; D_1(\theta,T)\,\Theta(\theta,\beta\varepsilon)
\;=\; D_1(\theta,T)\;\frac{1-\eta}{(1-\theta)(1+\eta)} ,
\label{eq:Dc}
\end{equation}
the Darken relation \cite{Darken1948,AlaNissila2002}: the same $\eta$ that shapes the static structure sets the ratio of chemical to jump diffusion.

Equation \eqref{eq:Dc} is the coefficient carried by the closed form
\eqref{eq:main}, whose rate is the local one $\Omega(\mathbf{K})$. A measurement
made at long times returns instead the coefficient of \eqref{eq:Dhydro}, smaller
by the collective correlation factor $R(\mathbf{0})=\Omega\tau_c$ of Sec.~\ref{sec:modes},
\begin{equation}
D_c^{\rm meas}(\theta,T) \;=\; \frac{D_1(\theta,T)\,\Theta(\theta,\beta\varepsilon)}
{R(\mathbf{0};\theta,\beta\varepsilon)} ,
\qquad R(\mathbf{0})\ \ge\ 1 ,
\label{eq:DcR}
\end{equation}
which is the correlation between successive moves that the mean-rate treatment
sets to unity. The two coincide wherever the relaxation proceeds through a single
mode, and in particular at $\varepsilon=0$ for every coverage, by \eqref{eq:sep};
away from that limit $R(\mathbf{0})$ is a property of the relaxation
spectrum of Sec.~\ref{sec:modes}. The thermodynamic factor is unaffected,
so that the coverage dependence contributed by $\Theta$ stands as written.

Had the linewidth instead been taken as $\Gamma_{\rm eff}[1-f_{\mathbf{K}}]$,
the result would have been $D_c=D_1$ and no Darken relation at all: it is the
factor $1/F$ forced onto the linewidth as $\mathbf{K}\to0$ that produces it,
in the form Darken's relation takes for a lattice
gas~\cite{AlaNissila2002}. On the surface the same relation holds with the
random-phase form \eqref{eq:FK2d}, giving
\begin{equation}
\Theta_{\rm RPA}(\theta,\beta\varepsilon)\;=\;\frac{1-\beta\varepsilon\,z\,\theta(1-\theta)}{1-\theta} ,
\label{eq:ThetaRPA}
\end{equation}
which does not reduce to the exact thermodynamic factor of \eqref{eq:Dc} at $z=2$, agreeing with it only to leading order in $\beta\varepsilon$, through $\eta\simeq\theta(1-\theta)\beta\varepsilon$: at $\theta=0.5$, $\beta\varepsilon=-1.5$ the exact factor is $4.23$ against $3.50$ from \eqref{eq:ThetaRPA}. The difference is the mean-field overestimate of the correlations, already met in Sec.~\ref{sec:ssf}.

\begin{figure}[!t]
\centering
\includegraphics[width=0.82\textwidth]{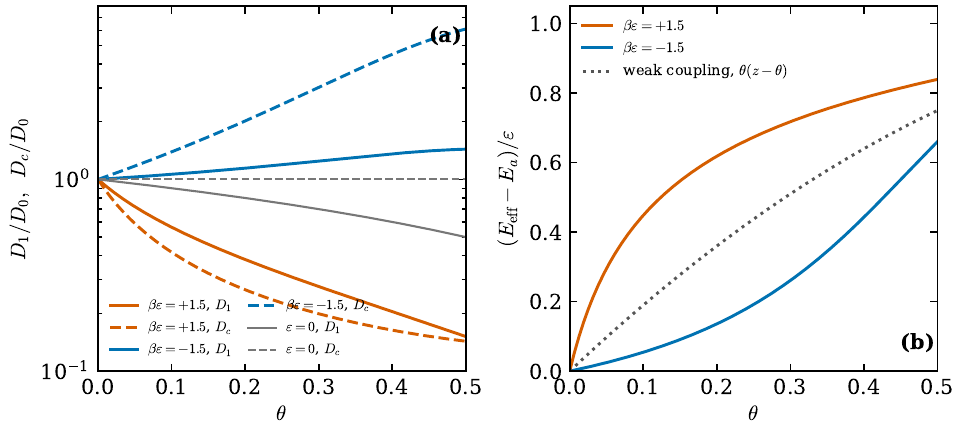}
\caption{The two faces of interacting transport along a channel ($z=2$), for the same attractive ($\beta\varepsilon=+1.5$, orange) and repulsive ($\beta\varepsilon=-1.5$, blue) adlayers in both panels, set throughout by $\eta(\theta,\beta\varepsilon)$. (a) The coefficients of Sec.~\ref{sec:transport}: jump $D_1$ (solid) and
chemical $D_c$ (dashed), Eqs.~\eqref{eq:Ds} and \eqref{eq:Dc}, normalized to $D_0$, with $\varepsilon=0$ in grey. Attraction suppresses both; repulsion enhances them, most strongly the chemical one, which rises about six-fold by $\theta=0.5$, while for $\varepsilon=0$ it stays exactly at $D_0$ \cite{Kutner1981}. (b) Their Arrhenius slope, Sec.~\ref{sec:arrhenius}: the excess
$(E_{\rm eff}-E_a)/\varepsilon$, Eq.~\eqref{eq:Eeff}, with the weak-coupling form $\theta(z-\theta)$ (dotted). The two panels are the magnitude and the temperature slope of the same interacting diffusion: attraction lowers the coefficients and raises the barrier, repulsion does the reverse.}
\label{fig:transport}
\end{figure}

Which of these coefficients an experiment actually returns must be stated. HeSE is fully coherent, so the observables are collective: the amplitude $F(\mathbf{K})$, the linewidth $\alpha(\mathbf{K})$, and the chemical coefficient $D_c$ defined by the $\mathbf{K}\to0$ limit of \eqref{eq:main}. The one-adsorbate expression \eqref{eq:isf-one} is an internal ingredient of
the theory, and $D_1$ of \eqref{eq:Ds} is a jump coefficient. The tracer coefficient belongs instead to an incoherent measurement and needs the correlation factor $f_c$; it should be assembled as $D_{\rm tr}=f_c D_1$, with $f_c$ from the literature \cite{AlaNissila2002}.
Both coefficients are elementary functions of $\theta$ and $\beta\varepsilon$ through $\eta$ and move oppositely with the sign of $\varepsilon$: attraction suppresses $D_c$ below $D_1/(1-\theta)$, repulsion raises it (Fig.~\ref{fig:transport}a). This turns the empirical blocking function of the ISA description \cite{TorresMiyares2026c} into a quantity predicted from $(\theta,\beta\varepsilon)$. The limits $\eta\to\pm1$ delimit the theory from within: as $\eta\to1$ $\Theta\to0$ and chemical diffusion shuts down; as $\eta\to-1$ $\Theta$ diverges and the layer becomes locally incompressible. In both the correlations grow strong and the closed forms lose accuracy.

\subsection{Effective activation energy}
\label{sec:arrhenius}

Since $D_1\propto\Gamma_{\rm eff}$ and the projection factor carries no
temperature, the Arrhenius slope of the jump coefficient at fixed coverage is $-\partial\ln\Gamma_{\rm eff}/\partial\beta$, an effective activation energy. The bare rate takes the transition-state form $\Gamma_0=\nu_0\,e^{-\beta E_a}$, with $E_a$ the saddle-point barrier and $\nu_0$ the attempt frequency \cite{HanggiTalknerBorkovec1990}; every other factor in $\Gamma_{\rm eff}$ depends on temperature only through $\beta\varepsilon$, and differentiating them gives the interaction correction in closed form,
\begin{equation}
E_{\rm eff}(\theta,T) \;\equiv\; -\,\frac{\partial \ln \Gamma_{\rm eff}}{\partial\beta}\bigg|_{\theta}
\;=\; E_a \;-\; \frac{\partial}{\partial\beta}\,
\ln\!\Big\{(1-\eta)\,\big[1-q+q\,e^{-\beta\varepsilon}\big]^{z-1}\Big\},
\label{eq:Eeff}
\end{equation}
the remaining derivative being elementary since $q=\theta+(1-\theta)\eta$ and $\partial\eta/\partial\beta=4\theta(1-\theta)\,\varepsilon\, e^{\beta\varepsilon}/[\zeta(\zeta+1)^{2}]$, with $\zeta$ from Eq.~\eqref{eq:eta}. The content is clearest at weak coupling, where $\eta\simeq\theta(1-\theta)\beta\varepsilon$ and $q\simeq\theta$: the correlation factor contributes $\theta(1-\theta)\varepsilon$ and the bond-counting factor $(z-1)\theta\varepsilon$, summing to
\begin{equation}
E_{\rm eff}(\theta) \;\simeq\; E_a \;+\; \varepsilon\,\theta\,(z-\theta),
\label{eq:Eeff-weak}
\end{equation}
both terms counting only initial-state bonds. Attraction therefore raises the barrier linearly in coverage, repulsion lowers it, and the slope measures $z\varepsilon$ directly. In Fig.~\ref{fig:transport}b the exact shift lies above the weak-coupling curve for attraction and below it for repulsion, because $\eta$ grows in magnitude with $\beta$. Since the ordinate there is divided by $\varepsilon$, this means that the exact treatment amplifies the attractive shift relative to weak coupling and moderates the repulsive one: at $\theta=0.5$ the three values are $0.84$, $0.75$ and $0.66$. The chemical channel carries a further, thermodynamic temperature dependence through $\Theta$, so Arrhenius plots of $D_1$ and $D_c$ at the same coverage need not share a slope.

\subsection{Confrontation with experiment and predictions}
\label{sec:predictions}

The closed forms make three predictions that a HeSE experiment can check directly. All are obtained from the standard fit $I=A+B\,e^{-\alpha t}$
[Eq.~\eqref{eq:diffusive}], so that no reinterpretation of the data is
required, and none is tied to a phase transition: they are ordinary, non-critical fingerprints of the lateral interaction.
(i) The diffusive amplitude is the structure factor, $B(\mathbf{K})=F(\mathbf{K})$: flat at $1-\theta$ for an ideal adlayer, gathered toward the zone centre for attraction and toward the zone boundary for repulsion [Fig.~\ref{fig:main}(a)].
(ii) The linewidth is narrowed in exact anticorrelation with that amplitude, so that the product $\alpha(\mathbf{K})F(\mathbf{K})=\Gamma_{\rm eff}[1-f_{\mathbf{K}}]$ carries no structure at all and varies only through the single-jump geometry [Fig.~\ref{fig:main}(c)]. This is the sharpest of the three, and it is a theorem: \eqref{eq:sumrule}
fixes the product without ever using the form of $F(\mathbf{K})$, so the
anticorrelation survives even where the Ising form does not. What is predicted is the momentum dependence of the product, $1-f_{\mathbf{K}}$; its magnitude is $\Gamma_{\rm eff}$ and carries the closures.
(iii) The Arrhenius slope shifts with coverage by $\varepsilon\,\theta(z-\theta)$, with a sign fixed by that of the interaction as shown in Fig.~\ref{fig:transport}(b).
Measured together, these three overdetermine $(\nu_0,E_a,\varepsilon)$, which turns the theory into an internal consistency test instead of a
one-parameter fit. One caveat attaches to that correspondence. Where the decay is not a single
exponential the fitted rate depends on the window, and the resulting bias is a
smooth multiplicative factor: it cancels from (i) and (ii), which rest on peak
positions and on a ratio, but not from absolute magnitudes, so it propagates
into $\nu_0$ and into (iii). Its sign is fixed, not uncertain: by
\eqref{eq:inst} the instantaneous rate falls monotonically, so a wider window
returns a smaller rate, by an amount governed by $\Delta(\mathbf{K})$. The
window must therefore be stated, and held fixed across a series. The
structural pair (i) and (ii) is already testable against published data; the
transport prediction (iii) has no equally clean data set, and is illustrated
instead on a layer whose interaction is roughly known.

\subsubsection{The repulsive adlayer: Na/Cu(111)}
\label{sec:nacu}

The cleanest existing test of the structural pair, (i) and (ii), is the dilute, strongly repulsive adlayer Na/Cu(111), for which coherent spin-echo data along the $[11\bar2]$ azimuth at $155$~K already exist \cite{Ward2021}, reporting two numbers at each momentum: the diffusive amplitude and the decay rate of the fit \eqref{eq:diffusive}.
And the system is chosen precisely because the short-range lattice-gas description fails on it: Na/Cu(111) is a strongly correlated, long-ranged dipolar layer whose structure factor is not the Ising one, its packing peak being a correlation-hole feature that no nearest-neighbour model would place. 
%
\begin{figure}[!t]
\centering
\includegraphics[width=0.99\textwidth]{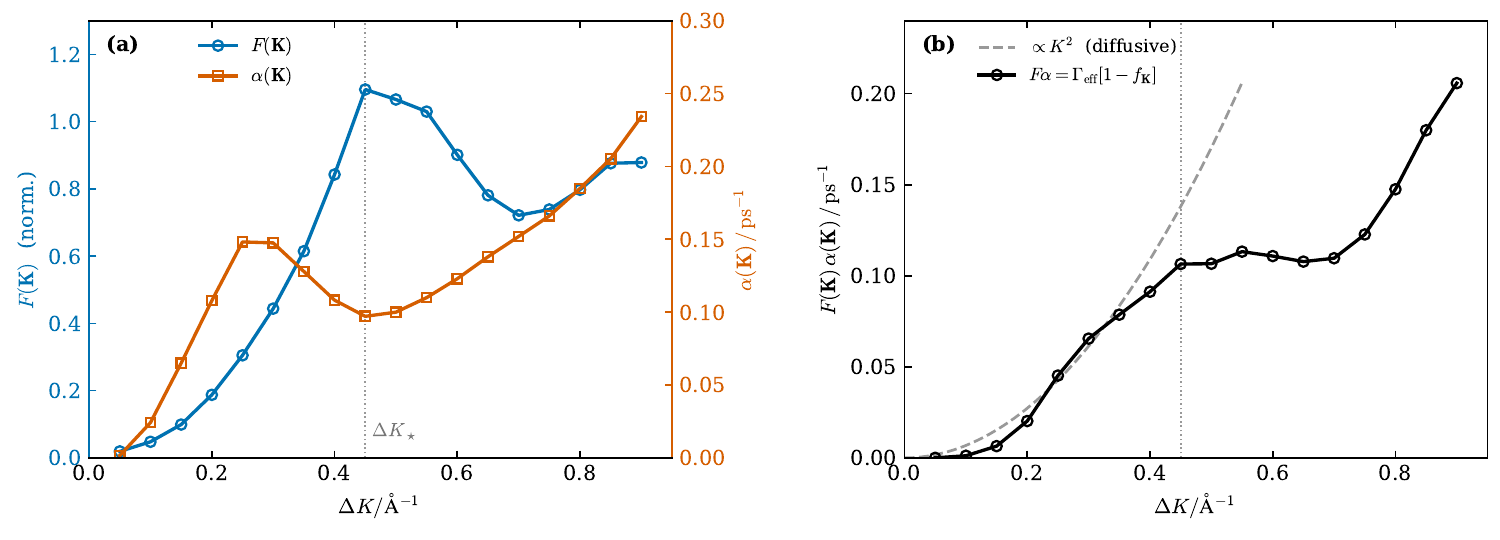}
\caption{Coherent HeSE data for Na/Cu(111) at $\theta=0.025$ ($[11\bar2]$ azimuth, $155$~K), digitized from Ref.~\cite{Ward2021} (Figs.~3b and 4a).
(a) The structural pair: the measured diffusive amplitude, our $F(\mathbf{K})$ (left axis), is suppressed as $\mathbf{K}\to0$ and peaks at $\Delta K^\star\approx0.45~\text{\AA}^{-1}$, while the decay rate $\alpha(\mathbf{K})$ (right axis) shows its anticorrelated dip at the same momentum: de Gennes narrowing, Eq.~\eqref{eq:alpha}.
(b) The parameter-free test: their point-by-point product $F\alpha=\Gamma_{\rm eff}[1-f_{\mathbf{K}}]$ loses all but a small residual of the peak and the dip and reduces to the smooth, diffusive single-adsorbate rate; equivalently, their ratio is $1/F(\mathbf{K})$, the narrowing itself.}
\label{fig:nacu}
\end{figure}
Take the amplitude first, Fig.~\ref{fig:nacu}(a). A repulsive layer keeps its adsorbates apart, so it carries almost no density
fluctuation at long wavelength: $F(\mathbf{K}\!\to\!0)\to0$. By
\eqref{eq:compress} that suppression is a statement of thermodynamics as much
as of geometry, $F(\mathbf{0})=\Theta^{-1}$ being the compressibility of the layer, which
repulsion lowers. The diffusive weight piles up instead at the momentum matching the typical
spacing $d$, a packing peak at $K^\star\simeq2\pi/d$. The measured amplitude does precisely this, peaking at $\Delta K^\star\approx0.45~\text{\AA}^{-1}$ against the geometric estimate $2\pi/d\approx0.42~\text{\AA}^{-1}$ for $d\approx15~\text{\AA}$. Here $d=n^{-1/2}$ with $n=\theta\times0.1768$~\AA$^{-2}$ the adsorbate density on Cu(111), which gives $15.0$~\AA\ at $\theta=0.025$. No interaction model enters, only the mean density.
Now the linewidth, in the same panel. Since $\alpha=\Gamma_{\rm eff}[1-f_{\mathbf{K}}]/F$, the structure factor that sits in the numerator of the amplitude sits in the denominator of the rate: wherever $F$ peaks, $\alpha$ must dip, and at the same momentum. The measured rate shows that dip at $\Delta K^\star$ --- de Gennes narrowing made visible.

The two together give a test that needs no fitting whatever as shown by Fig.~\ref{fig:nacu}(b). When the measured amplitude and rate are multiplied point by point, the shared $F$ drops out because $F\alpha=\Gamma_{\rm eff}[1-f_{\mathbf{K}}]$: peak and dip, each a factor $\approx1.5$, largely cancel, and two strongly structured curves collapse onto one smooth, diffusive ($\propto K^{2}$) single-adsorbate rate, leaving only a residual shoulder between about $0.5$ and
$0.7~\text{\AA}^{-1}$ an order of magnitude smaller than either feature, and
no free parameter enters. The original analysis adopted this factorization as
an assumption \cite{Ward2021}; here it follows from a single construction.
\begin{figure}[!t]
\centering
\includegraphics[width=0.99\textwidth]{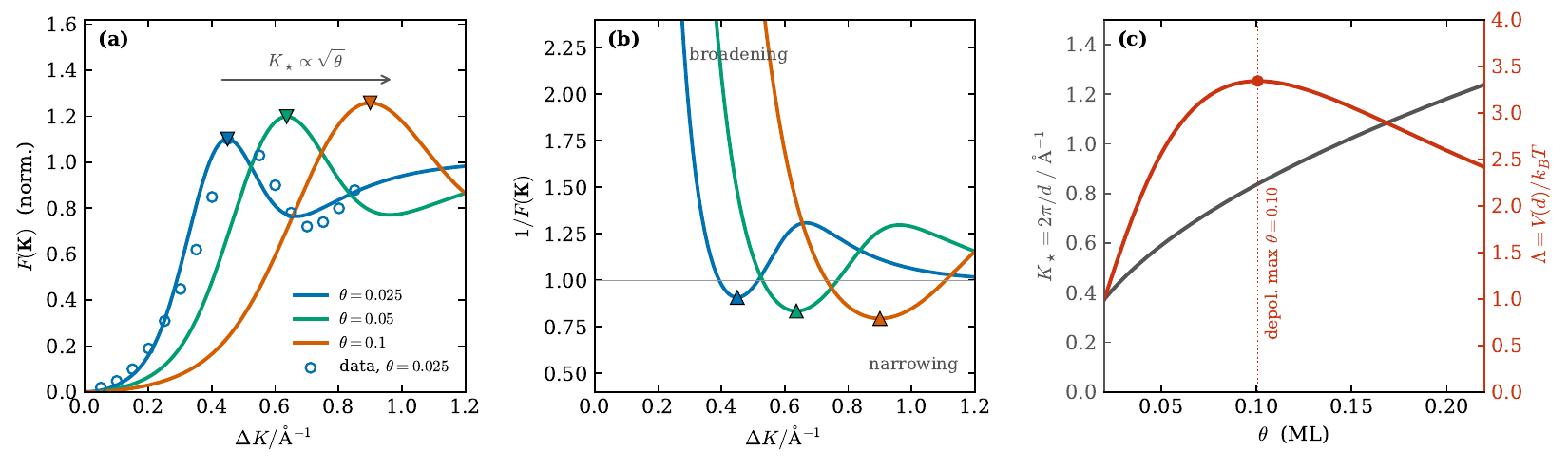}
\caption{Prediction for Na/Cu(111) at raised coverage, combining the geometric shift of the packing peak with Topping depolarization from the measured Kohn--Lau parameters \cite{Rittmeyer2016}; the $\theta=0.025$ points of Fig.~\ref{fig:nacu} are repeated as open circles.
(a) The amplitude at $\theta=0.025$, $0.05$ and $0.10$, with the correlation peak marked by triangles at $K^\star(\theta)$ of \eqref{eq:kstar}. (b) The corresponding narrowing factor $1/F(\mathbf{K})$, with the compensating broadening in the flanks. The profiles in (a) and (b) are schematic, no closed form being available at this coupling (see text): the shape is fitted once to the measured $\theta=0.025$ amplitudes, every length then scales as $\sqrt{\theta}$, the peak contrast is scaled by $\Lambda(\theta)$, and each maximum is locked to $K^\star(\theta)$.
(c) The two behaviours separated, and this panel is quantitative and free of adjustable parameters: the peak position (grey) marches monotonically as $2\pi/d(\theta)$, whereas the dipolar coupling $\Lambda(\theta)$ (red) turns over at $\theta=0.10$, where depolarization overtakes the density gain.}
\label{fig:nacupred}
\end{figure}
Raising the coverage separates two effects that the single data set cannot. The first is geometric. The mean spacing shrinks as $d\propto\theta^{-1/2}$, so the packing peak, and the narrowing dip riding on it, must march to higher momentum as
\begin{equation}
K^\star(\theta)\;\simeq\;K^\star(\theta_0)\,\sqrt{\theta/\theta_0},
\label{eq:kstar}
\end{equation}
from $\approx0.45~\text{\AA}^{-1}$ at $\theta_0=0.025$ to $\approx0.64$ and $0.90~\text{\AA}^{-1}$ at $\theta=0.05$ and $0.10$ as plotted in Figs.~\ref{fig:nacupred}(a,b). The shift is set by the density alone.
The second is not. The depth of the narrowing is fixed by the interaction strength, and it need not grow with coverage, because the dipoles depolarize as they crowd. With the measured Kohn--Lau dipole moment and its Topping depolarization \cite{Topping1927,Rittmeyer2016}, the effective moment falls from $\mu_{\rm eff}\approx6.9$~D at $\theta=0.025$ to $3.9$~D at $\theta=0.10$, so the dipolar coupling $\Lambda(\theta)=V(d)/k_BT$ that sets the contrast first rises ($1.3\to2.6\to3.4$) and then turns over near $\theta\approx0.10$ (see Fig.~\ref{fig:nacupred}(c)). Here $V(d)=2\mu_{\rm eff}^{2}/4\pi\epsilon_0 d^{3}$ is the dipolar repulsion at the mean spacing, and the depolarization follows $\mu_{\rm eff}=\mu_0/(1+9\alpha n^{3/2})$ with $\mu_0\simeq7.8$~D and $\alpha\simeq47$~\AA$^{3}$. The joint prediction
is distinctive, and that is why it repays measurement: both behaviours are
accessible on existing apparatus.

The line shape itself is another matter. The collective form \eqref{eq:FKrpa} holds for an interaction of any range, and one might expect to use it here with the dipolar $v(\mathbf{K})$; but evaluated with the hard-core-truncated dipolar repulsion it reaches its spinodal already at $\Lambda\simeq2.6$, that is at $\theta\approx0.05$, so over most of the window studied it returns a divergence instead of a profile. A strongly correlated dipolar liquid is outside the reach of a Gaussian
closure. Quantitative here are therefore the peak positions and the coupling
$\Lambda(\theta)$, which come from \eqref{eq:kstar} and from the Topping depolarization; the profiles of Fig.~\ref{fig:nacupred}(a,b) are schematic,
their shape taken where needed from ab initio molecular
dynamics~\cite{Rittmeyer2016}, in which the de Gennes feature appears only
once the dipole interaction is included. And above $\theta\approx0.25$ the layer becomes commensurately ordered, so the window $0.025\lesssim\theta\lesssim0.15$ studied here stays clear of it.

\subsubsection{The attractive adlayer: H/Pt(111)}
\label{sec:hpt}

Where Sec.~\ref{sec:nacu} tested only what survives whatever the SSF may be,
the present section uses the closed forms \eqref{eq:FK}, \eqref{eq:FK2d} and
\eqref{eq:gamma-eff} to turn coverage and interaction strength into numbers,
and is therefore only as good as the short-range description behind them: an
illustration carrying an explicit range on $\varepsilon$, not a fit. The theory is best seen on an attractive adlayer, where the signatures are largest and gather at the zone centre, and hydrogen on Pt(111) is the natural choice: a triangular lattice ($z=6$), an attractive NN interaction, and the system worked out in most detail within the CF framework \cite{TorresMiyares2026c}.

Density-functional estimates of the H--H interaction on Pt(111) scatter in both sign and size, so instead of pinning a single number we carry the range $\varepsilon\approx12$--$20$~meV attractive, around a representative $16$~meV \cite{Yu2018}; were the interaction repulsive instead, the construction would be identical with $\eta<0$ and the signatures moved to the zone boundary. We work at $\theta=0.2$ and $T=250$~K throughout.
An attraction of strength $\varepsilon$ carries an ordering temperature,
$\varepsilon/(k_B\ln 3)$ for the triangular lattice at half
coverage~\cite{Baxter1982}, here $127$--$211$~K; below it the layer condenses into islands and the uniform-layer description fails. Since a layer at $\theta=0.2$ orders below the half-coverage value, this is an
upper bound: staying above it at $T=250$~K across the whole range is a sufficient condition, not a marginal one. The dimensionless coupling is then $\beta\varepsilon\approx0.56$--$0.93$, which \eqref{eq:eta} turns into $\eta\approx0.10$--$0.17$. Every number that follows is a function of that one.

\begin{figure}[!t]
\centering
\includegraphics[width=0.82\textwidth]{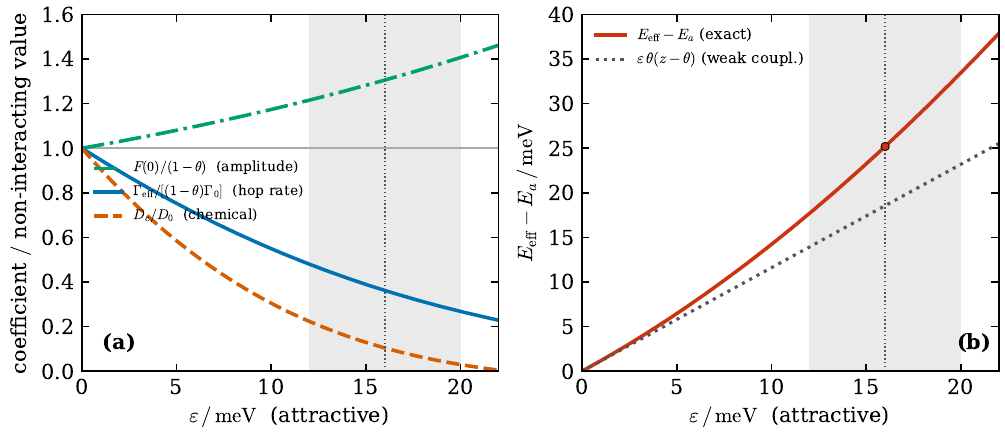}
\caption{Correlated response of the H/Pt(111) coefficients to the one uncertain input, the interaction strength, at fixed $\theta=0.2$ ($z=6$, $T=250$~K); complementary to Fig.~\ref{fig:transport}, which fixes the interaction and varies the coverage instead. The shaded band marks the density-functional range $\varepsilon\approx12$--$20$~meV, the dotted vertical line the representative $16$~meV.
(a) Magnitudes, each relative to its $\varepsilon=0$ value: the hop rate $\Gamma_{\rm eff}/[(1-\theta)\Gamma_0]$ --- equivalently the jump coefficient $D_1/[(1-\theta)D_0]$ --- and the chemical coefficient $D_c/D_0$, built with $\Theta_{\rm RPA}$ of \eqref{eq:ThetaRPA}, are suppressed, while the zone-centre amplitude $F(\mathbf{0})/(1-\theta)$, drawn here with the pair-approximation form $(1+\eta)/(1-\eta)$, that is, the lower edge of the bracket discussed in the text, is enhanced; all three move away from unity together, with $D_c$ bending toward zero as the collective closure approaches its mean-field spinodal.
(b) Temperature slope: the excess $E_{\rm eff}-E_a$ of Eq.~\eqref{eq:Eeff} above the weak-coupling estimate $\varepsilon\,\theta(z-\theta)$ (dotted). All four quantities are elementary functions of $(\theta,\beta\varepsilon)$.}
\label{fig:hpt}
\end{figure}

The hop rate is suppressed twice over \eqref{eq:gamma-eff}. Attraction makes the target site harder to find empty, and it adds bonds that must be broken before the adsorbate can leave. Together these drop $\Gamma_{\rm eff}$ to $(0.38$--$0.21)\,\Gamma_0$, below even the bare site-blocking value $(1-\theta)\,\Gamma_0=0.8\,\Gamma_0$ --- a suppression a spin-echo measurement resolves comfortably. The jump coefficient goes with it, since $D_1=\avg{\ell^{2}}\,\Gamma_{\rm eff}$ by \eqref{eq:Ds} falls by exactly the same factor, Fig.~\ref{fig:hpt}(a).

The collective coefficient falls further still. It carries in addition the thermodynamic factor $\Theta_{\rm RPA}$ of \eqref{eq:ThetaRPA}, which attraction drives well below unity, $\Theta\approx0.58$--$0.14$, so that $D_c=D_1\Theta$ collapses to $(0.22$--$0.03)\,D_0$. Attraction thus slows collective diffusion far more than it slows an individual hop --- the clustering that the Darken relation encodes. The strong end of the range needs a caution. A vanishing $\Theta$ is a vanishing $\sigma(\mathbf{0})$ in
\eqref{eq:stiffness}: attraction has cancelled the entropic term $k_BT/S_0$, the uniform layer
no longer resists a long-wavelength modulation of its density, and it would
separate into dense and dilute regions --- the island condensation already met
in the ordering temperature above. Mean field places that instability at
$\beta\varepsilon\approx1.04$ for $\theta=0.2$, and our largest interaction,
$\beta\varepsilon\approx0.93$, sits close to it, where the closure
overestimates the correlations. The smallest values, $\Theta\approx0.14$ and $D_c\approx0.03\,D_0$, are therefore a trend, not a prediction.

The temperature slope moves with them. At $\theta=0.2$ the effective activation energy \eqref{eq:Eeff} exceeds the bare barrier by $\Delta E_{\rm eff}\approx18$--$33$~meV, above the weak-coupling estimate $\varepsilon\,\theta(z-\theta)\approx14$--$23$~meV because the correlations stiffen as the temperature drops [Fig.~\ref{fig:hpt}(b)]. The excess sits well within the resolution of a HeSE Arrhenius plot.

The structural signature completes the set. For an attractive layer the diffusive amplitude gathers toward the zone centre, $F(\mathbf{0})>1-\theta$.
Its size is the one quantity the mean-field statics do not fix cleanly: the pair approximation gives $F(\mathbf{0})/(1-\theta)=(1+\eta)/(1-\eta)\approx1.2$--$1.4$ and the collective form \eqref{eq:FK2d} gives $2.1$--$9.2$ across the same range, so the two only bracket the enhancement, the spread widening as that temperature is approached. Even the lower bound of the bracket, a $20$--$40\%$ enhancement, is within
reach of a spin-echo amplitude measurement.


The correlated motion is the test. Vary $\varepsilon$, $\theta$ or $T$, and the hop rate, both diffusion coefficients, the activation energy and the amplitude all move together, because all five are functions of $\eta(\theta,\beta\varepsilon)$ alone.

\section{Concluding remarks}
\label{sec:concl}

The ISF of an interacting adlayer, and with it the whole hierarchy of
linear-response functions, follows in closed form once the lateral interaction is
carried into the problem through the single correlation parameter
$\eta(\theta,\beta\varepsilon)$. The construction rests on the double role of the
ISF as a CF, of the adsorbate displacements in time and of
the adsorbate separations at $t=0$, and the closed form \eqref{eq:main} is an
elementary function of $\mathbf{K}$, $t$, coverage and temperature. What
distinguishes the present treatment is that the form is derived, not assembled. Eliminating every variable of the layer but the separation leaves an exact
equation of motion, \eqref{eq:eom}. Its amplitude is the identity
$I(\mathbf{K},0)=F(\mathbf{K})$, and its local rate obeys the sum rule
\eqref{eq:sumrule}, in which the interaction enters only through the scalar
exchange current \eqref{eq:current}. Discarding the memory function returns
\eqref{eq:main}. The narrowing of the line by the inverse
structure factor is therefore a theorem and the residual error is bounded.

Two approximations enter, and both are static. The SSF is exact in
one dimension by the transfer matrix, but on the surface it rests on the
random-phase treatment, which overestimates the growth of correlations. 
The hop rate \eqref{eq:gamma-eff} takes its two averages independently, which
is exact along a channel and a pair approximation on the surface. Neither is
tested here, and neither concerns the dynamics: both are equilibrium
closures, and measuring how far they depart from the truth on a surface is a
numerical question, which a companion study of the same Hamiltonian under
Kawasaki dynamics will address. What the dynamics contributes is not an approximation of uncontrolled sign but
a truncation whose sign is fixed. The memory function being non-negative,
\eqref{eq:main} is a rigorous lower bound on the ISF at every momentum
transfer and at every time, \eqref{eq:bound}. Its accuracy is fixed by a
single static number, the spread $\Delta(\mathbf{K})$ of the mode rates about
their mean, which vanishes identically at $\varepsilon=0$, where the density wave is an exact relaxation mode by
\eqref{eq:sep}, which is why the symmetric exclusion process is returned exactly, and not by coincidence, and it grows only as the correlations build.

Two statements of the theory are of unequal robustness, and the confrontation
with data depends on keeping them apart. That the product
$\alpha(\mathbf{K})F(\mathbf{K})=\Gamma_{\rm eff}[1-f_{\mathbf{K}}]$ carries no
structure follows from detailed balance and from the moves being exchanges
between a fixed set of neighbour vectors, 
so that it
survives where the Ising description does not. Predicting $F$ and
$\Gamma_{\rm eff}$ themselves from $(\theta,\varepsilon)$ through $\eta$ is a
different matter, and does require the interaction to be short-ranged or else
supplied externally as a $v(\mathbf{K})$. Section~\ref{sec:nacu} exploits the
split: Na/Cu(111) is a long-ranged, strongly correlated dipolar layer on which
the lattice-gas closed form has no claim, and the cancellation is confirmed there
all the same.

How far the second can be trusted depends on coverage. The error of the
surface closures is set by the correlation strength, largest at half coverage
through $\theta(1-\theta)$; they are least reliable near $\theta=\tfrac12$ and
safest in the dilute and near-saturation wings. From within the theory,
$|\eta|\to1$ marks the strong-coupling limit beyond which the closed forms
lose accuracy. That rule is specific to short-ranged
interactions: a long-ranged repulsion can remain strongly correlated even when
dilute, and there $F$ must be taken from the packing geometry, from simulation or
from the measurement itself, while the structureless product survives intact.
The accuracy of the decay shape is governed separately, and quantitatively, by
$\Delta(\mathbf{K})$: by \eqref{eq:excess} the closed form is within ten per cent of the exact
signal at $\Omega t=1$ whenever $\Delta$ stays below about one fifth.

One objection arises from that division.
If the SSF is to be taken from the measurement
wherever the layer is strongly correlated, then the relation that survives is
being tested only where it can hardly fail, while the part of the theory that
genuinely predicts is exercised only where it is safe. The objection is fair, and
the reply is not that it is wrong but that it is bounded, in three ways. The product relation is not empty. It would retain structure if the coherent
relaxation were not carried by a single density mode, if the jump geometry
departed from the assumed neighbour shell, or if the amplitude returned by the
fit were not the layer's own structure factor. On the one data set available
it does not. The predictive statements are exposed, not sheltered: Sec.~\ref{sec:hpt}
commits to a set of coefficients across an explicit range of $\varepsilon$,
any one of which a measurement can contradict. And the boundary
between the two is drawn in advance, by $\theta(1-\theta)$, by $|\eta|$, by the
spinodal of the collective closure and now by $\Delta(\mathbf{K})$, before any comparison is made and not after
it.

The memory function that the closed form discards is also an observable, and that
is the direction in which the treatment is most readily extended. Two of its
consequences require no fitting at all: the height of the quasi-elastic peak
relative to the closed-form Lorentzian returns $\Omega\tau_c$ directly, and the
rate obtained from a spin-echo fit must fall as the fitting window is widened, by
an amount that tracks $\Delta(\mathbf{K})$ across the zone. Measuring either
turns the width of the relaxation spectrum into a number, and with it the
collective correlation factor $R(\mathbf{0})$ that separates the chemical
diffusion coefficient of \eqref{eq:DcR} from the mean-rate one. Three further directions are open. Longer-ranged and even oscillatory
substrate-mediated interactions enter the collective form only through
$v(\mathbf{K})$ in \eqref{eq:FKrpa}, which puts the Friedel-oscillation tails
of adsorbates on metals formally within reach. The mean-field statics can be
improved by higher levels of the cluster-variation method
\cite{Pelizzola2005}, or $F$ supplied from simulation or from the measurement,
without touching the rest of the construction. And substrate friction and
quantum corrections
\cite{TorresMiyares2026a,TorresMiyares2026b,TorresMiyares2026c} act only through
$\Gamma_0$, so that the interacting theory inherits them unchanged. One quantity
the theory does not reach at all: the tracer correlation factor $f_c$, which a
coherent experiment does not report and which Sec.~\ref{sec:transport} therefore
takes from the literature instead of predicting. Closing that gap would require
following a tagged adsorbate, and is a separate problem from the one solved here.

The upshot is a closed-form link between the lateral interaction and the coherent
spin-echo signal, obtained as the first truncation of an exact equation of motion, not as a
model. Sign, strength and range are written into the diffusive
amplitude, into its linewidth and into the Arrhenius slope, all three organized by the single function $\eta(\theta,\beta\varepsilon)$.
The theory predicts not three numbers but their correlated motion, since all of them are elementary functions of
$(\theta,\beta\varepsilon)$. The accuracy of the prediction is itself controlled, by a static width whose vanishing marks the one layer for which the
closed form is not an approximation at all.


\begin{thebibliography}{99}
\setlength{\itemsep}{0pt}
\bibitem{Jardine2009}
A.~P.~Jardine, G.~Alexandrowicz, H.~Hedgeland, W.~Allison, and J.~Ellis,
\emph{Studying the microscopic nature of diffusion with helium-3 spin-echo},
Phys.\ Chem.\ Chem.\ Phys.\ \textbf{11}, 3355 (2009).

\bibitem{TorresMiyares2026a}
E.~E.~Torres-Miyares and S.~Miret-Art\'es,
\emph{The characteristic function method in surface diffusion},
Phys.\ Chem.\ Chem.\ Phys.\ \textbf{28}, 8916 (2026).

\bibitem{TorresMiyares2026b}
E.~E.~Torres-Miyares and S.~Miret-Art\'es,
\emph{Surface diffusion at finite coverage: the characteristic function method},
Surfaces \textbf{9}, 32 (2026).

\bibitem{TorresMiyares2026c}
E.~E.~Torres-Miyares and S.~Miret-Art\'es,
\emph{The characteristic function as a unifying framework for linear response in surface diffusion},
Surfaces \textbf{9}, 72 (2026).

\bibitem{Montroll1965}
E.~W.~Montroll and G.~H.~Weiss,
\emph{Random walks on lattices. II},
J.\ Math.\ Phys.\ \textbf{6}, 167 (1965).

\bibitem{Chudley1961}
C.~T.~Chudley and R.~J.~Elliott,
\emph{Neutron scattering from a liquid on a jump diffusion model},
Proc.\ Phys.\ Soc.\ \textbf{77}, 353 (1961).

\bibitem{Frenken1992}
J.~W.~M.~Frenken and B.~J.~Hinch,
\emph{Quasielastic helium scattering studies of adatom diffusion on surfaces},
in \emph{Helium Atom Scattering from Surfaces}, Springer (1992), pp.~287--313.

\bibitem{nakajima}
S. Nakajima, Prog. Theor. Phys. \textbf{20}, 948 (1958).

\bibitem{zwanzig60}
R. Zwanzig, J. Chem. Phys. \textbf{33}, 1338 (1960).

\bibitem{MartinezCasado2007}
R.~Mart\'inez-Casado, J.~L.~Vega, A.~S.~Sanz, and S.~Miret-Art\'es,
\emph{Line shape broadening in surface diffusion of interacting adsorbates with quasielastic He atom scattering},
Phys.\ Rev.\ Lett.\ \textbf{98}, 216102 (2007).

\bibitem{Ellis2001}
J.~Ellis, A.~P.~Graham, F.~Hofmann, and J.~P.~Toennies,
\emph{Coverage dependence of the microscopic diffusion of Na atoms on the Cu(001) surface: a combined helium atom scattering experiment and molecular dynamics study},
Phys.\ Rev.\ B \textbf{63}, 195408 (2001).

\bibitem{Alexandrowicz2006}
G.~Alexandrowicz, A.~P.~Jardine, H.~Hedgeland, W.~Allison, and J.~Ellis,
\emph{Onset of 3D collective surface diffusion in the presence of lateral interactions: Na/Cu(001)},
Phys.\ Rev.\ Lett.\ \textbf{97}, 156103 (2006).

\bibitem{Reed1981}
D.~A.~Reed and G.~Ehrlich,
\emph{Surface diffusion, atomic jump rates and thermodynamics},
Surf.\ Sci.\ \textbf{102}, 588 (1981).

\bibitem{Gomer1990}
R.~Gomer,
\emph{Diffusion of adsorbates on metal surfaces},
Rep.\ Prog.\ Phys.\ \textbf{53}, 917 (1990).

\bibitem{Uebing1991}
C.~Uebing and R.~Gomer,
\emph{A Monte Carlo study of surface diffusion coefficients in the presence of adsorbate--adsorbate interactions. I--IV},
J.\ Chem.\ Phys.\ \textbf{95}, 7626--7652 (1991).

\bibitem{AlaNissila2002}
T.~Ala-Nissila, R.~Ferrando, and S.~C.~Ying,
\emph{Collective and single particle diffusion on surfaces},
Adv.\ Phys.\ \textbf{51}, 949 (2002).

\bibitem{deGennes1959}
P.~G.~de Gennes,
\emph{Liquid dynamics and inelastic scattering of neutrons},
Physica \textbf{25}, 825 (1959).

\bibitem{Kutner1981}
R.~Kutner,
\emph{Chemical diffusion in the lattice gas of non-interacting particles},
Phys.\ Lett.\ A \textbf{81}, 239 (1981).

\bibitem{Ward2021}
D.~J.~Ward, A.~Raghavan, A.~Tamt\"ogl, A.~P.~Jardine, E.~Bahn, J.~Ellis, S.~Miret-Art\'es, and W.~Allison,
\emph{Inter-adsorbate forces and coherent scattering in helium spin-echo experiments},
Phys.\ Chem.\ Chem.\ Phys.\ \textbf{23}, 7799 (2021).

\bibitem{torres2026}
E. E. Torres-Miyares and S. Miret-Art\'es, Eur. Phys. J. Plus \textbf{141}, 723 (2026).

\bibitem{Mori1965}
H.~Mori,
\emph{Transport, collective motion, and Brownian motion},
Prog.\ Theor.\ Phys.\ \textbf{33}, 423 (1965).

\bibitem{Zwanzig2001}
R.~Zwanzig,
\emph{Nonequilibrium Statistical Mechanics}
(Oxford University Press, Oxford, 2001).

\bibitem{Kawasaki1966}
K.~Kawasaki,
\emph{Diffusion constants near the critical point for time-dependent Ising models. III},
Phys.\ Rev.\ \textbf{150}, 285 (1966).

\bibitem{liggett}
T. M. Liggett, Continuous Time Markov Processes: An Introduction, Graduate Studies in Mathematics \textbf{113} (American Mathematical Society, Providence, 2010).

\bibitem{Baxter1982}
R.~J.~Baxter,
\emph{Exactly Solved Models in Statistical Mechanics},
Academic Press (1982).

\bibitem{Pelizzola2005}
A.~Pelizzola,
\emph{Cluster variation method in statistical physics and probabilistic graphical models},
J.\ Phys.\ A: Math.\ Gen.\ \textbf{38}, R309 (2005).

\bibitem{Hansen2013}
J.~P.~Hansen and I.~R.~McDonald,
\emph{Theory of Simple Liquids: With Applications to Soft Matter}, 4th ed.,
Academic Press (2013).

\bibitem{ChaikinLubensky}
P.~M.~Chaikin and T.~C.~Lubensky,
\emph{Principles of Condensed Matter Physics},
Cambridge University Press (1995).

\bibitem{Repp2000}
J.~Repp, F.~Moresco, G.~Meyer, K.-H.~Rieder, P.~Hyldgaard, and M.~Persson,
\emph{Substrate mediated long-range oscillatory interaction between adatoms: Cu/Cu(111)},
Phys.\ Rev.\ Lett.\ \textbf{85}, 2981 (2000).

\bibitem{Darken1948}
L.~S.~Darken,
\emph{Diffusion, mobility and their interrelation through free energy in binary metallic systems},
Trans.\ AIME \textbf{175}, 184 (1948).

\bibitem{HanggiTalknerBorkovec1990}
P.~H\"anggi, P.~Talkner, and M.~Borkovec,
\emph{Reaction-rate theory: fifty years after Kramers},
Rev.\ Mod.\ Phys.\ \textbf{62}, 251 (1990).

\bibitem{Rittmeyer2016}
S.~P.~Rittmeyer, D.~J.~Ward, P.~G\"utlein, J.~Ellis, W.~Allison, and K.~Reuter,
\emph{Energy dissipation during diffusion at metal surfaces: disentangling the role of phonons versus electron--hole pairs},
Phys.\ Rev.\ Lett.\ \textbf{117}, 196001 (2016).

\bibitem{Topping1927}
J.~Topping,
\emph{On the mutual potential energy of a plane network of doublets},
Proc.\ R.\ Soc.\ Lond.\ A \textbf{114}, 67 (1927).

\bibitem{Yu2018}
C.~Yu, F.~Wang, Y.~Zhang, L.~Zhao, B.~Teng, M.~Fan, and X.~Liu,
\emph{H$_2$ thermal desorption spectra on Pt(111): a density functional theory and kinetic Monte Carlo simulation study},
Catalysts \textbf{8}, 450 (2018).

\end{thebibliography}
\end{document}